\documentclass[conference,9pt]{IEEEtran}
\IEEEoverridecommandlockouts
\usepackage{amsmath,amssymb,amsthm,amsfonts,bm}
\usepackage{graphicx}
\usepackage{booktabs,multirow,array}
\usepackage{xcolor}
\usepackage{tikz}
\usetikzlibrary{arrows.meta,positioning,fit,calc,shapes.misc,backgrounds,matrix,decorations.pathreplacing,patterns,patterns.meta}
\usetikzlibrary{shapes.geometric}

\definecolor{fzink}{HTML}{0B0B0B}     \definecolor{fzinkii}{HTML}{52514E}
\definecolor{fzref}{HTML}{898781}     \definecolor{fzgrid}{HTML}{E1E0D9}
\definecolor{fzours}{HTML}{2A78D6}    \definecolor{fzoursdk}{HTML}{1C5CAB}
\definecolor{fzvf}{HTML}{EB6834}      \definecolor{fzvfdk}{HTML}{B94E1F}
\definecolor{fzkA}{HTML}{6DA7EC}      \definecolor{fzkB}{HTML}{B7D3F6}   \definecolor{fzkC}{HTML}{DDEAF9}
\definecolor{fzport}{HTML}{D2D0CB}
\definecolor{fzdie}{HTML}{F4F2EE}     \definecolor{fzip}{HTML}{E3E1DB}   \definecolor{fzsub}{HTML}{EAE3D3}
\definecolor{fzbump}{HTML}{C3C0B9}    \definecolor{fztsv}{HTML}{8F8C85}
\definecolor{fzcu}{HTML}{E6CF86}      \definecolor{fzcudk}{HTML}{8E7442}
\definecolor{fzgnd}{HTML}{D3D0C8}
\definecolor{fzband}{HTML}{F3F2EF}    \definecolor{fzoband}{HTML}{EDF3FC}
\definecolor{fzmesh}{HTML}{A3A099}

\newcommand{\fzm}[1]{{\footnotesize$#1$}}
\newcommand{\fzs}[1]{#1}
\newcommand{\fzT}{^{\fzs{\mathsf{T}}}}

\tikzset{
  fz/.style={every node/.style={}, font=\sffamily\scriptsize, text=fzink, line join=round,
             >={Stealth[length=3.2pt,width=2.6pt]}},
  fzlbl/.style={font=\sffamily\scriptsize, text=fzink, inner sep=1pt, align=center},
  fznote/.style={font=\sffamily\scriptsize, text=fzinkii, inner sep=1pt, align=center},
  fzmath/.style={font=\footnotesize, text=fzink, inner sep=0pt},
  fztitle/.style={font=\sffamily\footnotesize, text=fzink, inner sep=1pt, align=center},
  fztag/.style={font=\sffamily\footnotesize\bfseries, text=#1, inner sep=1pt},
  fzcard/.style={draw=fzmesh, line width=0.5pt, fill=white, rounded corners=1.5pt},
  fzocard/.style={draw=fzours, line width=0.5pt, fill=white, rounded corners=1.5pt},
  fzflow/.style={->, draw=fzinkii, line width=0.6pt, rounded corners=2pt},
  fzoflow/.style={->, draw=fzoursdk, line width=0.8pt, rounded corners=2pt},
  fzprior/.style={draw=fzref, line width=0.5pt, dash pattern=on 2pt off 1.5pt, rounded corners=2.5pt},
  fzzoom/.style={draw=fzinkii, line width=0.4pt, dash pattern=on 1.5pt off 1pt},
  fznum/.style={circle, draw=fzinkii, line width=0.5pt, fill=white, text=fzink, font=\sffamily\bfseries\scriptsize,
                inner sep=0pt, minimum size=2.7mm},
  fzbadge/.style={rounded rectangle, draw=#1, line width=0.4pt, fill=#1!7, font=\sffamily\scriptsize,
                  text=fzink, inner xsep=2.2pt, inner ysep=0.9pt, minimum height=3.3mm, text height=2.25mm, text depth=0.5mm},
  fznport/.style={rectangle, fill=fzink, minimum size=1.35mm, inner sep=0pt},
  fzntrace/.style={circle, draw=fzoursdk, fill=white, line width=0.5pt, minimum size=1.5mm, inner sep=0pt},
  fznvia/.style={diamond, draw=fzoursdk, fill=fzours, line width=0.4pt, minimum size=2.1mm, inner sep=0pt},
  fznpad/.style={rectangle, rounded corners=0.35mm, draw=fzoursdk, fill=fzkB, line width=0.5pt,
                 minimum size=1.6mm, inner sep=0pt},
  fznref/.style={rectangle, rounded corners=0.3mm, draw=fzinkii, fill=fzgnd, line width=0.4pt,
                 minimum width=2.6mm, minimum height=1.25mm, inner sep=0pt},
  fzeser/.style={draw=fzinkii, line width=0.5pt},
  fzecpl/.style={draw=fzinkii, line width=0.4pt, dash pattern=on 1.2pt off 0.9pt},
  fzeref/.style={draw=fzinkii, line width=0.4pt, dash pattern=on 0.4pt off 0.8pt, line cap=round},
  pics/fzclock/.style={code={
    \draw[#1, line width=0.5pt, fill=white] (0,0) circle[radius=1.12mm];
    \draw[#1, line width=0.5pt, line cap=round] (0,0) -- (0,0.72mm) (0,0) -- (0.52mm,-0.3mm);}},
  pics/fzwarn/.style={code={
    \filldraw[#1, line width=0.5pt, line join=round, fill=white] (90:1.42mm) -- (210:1.42mm) -- (330:1.42mm) -- cycle;
    \draw[#1, line width=0.6pt, line cap=round] (0,0.40mm) -- (0,-0.18mm);
    \fill[#1] (0,-0.47mm) circle[radius=0.17mm];}},
  pics/fzcheck/.style={code={
    \draw[#1, line width=0.5pt, fill=white] (0,0) circle[radius=1.12mm];
    \draw[#1, line width=0.65pt, line cap=round, line join=round] (-0.52mm,0.02mm) -- (-0.14mm,-0.4mm) -- (0.55mm,0.45mm);}},
}
\newcommand{\fzicon}[1]{\tikz[baseline=-0.6ex]\pic{#1};\,}

\newcommand{\fzXsection}{%
  \filldraw[draw=fzinkii, line width=0.4pt, fill=fzsub] (0,0) rectangle (34,3.0);
  \foreach \x in {3.0,6.4,9.8,24.4,27.6,30.8} {\filldraw[draw=fzinkii, line width=0.4pt, fill=fzbump] (\x,3.85) circle[radius=0.82];}
  \filldraw[draw=fzinkii, line width=0.4pt, fill=fzip] (0.6,4.65) rectangle (33.4,9.0);
  \foreach \x in {24.4,27.6,30.8} {\fill[fztsv] (\x-0.33,4.65) rectangle (\x+0.33,8.15);}
  \draw[fztsv, line width=0.6pt] (29.0,9.0) -- (29.0,8.55) -- (27.6,8.55) -- (27.6,8.15);
  \foreach \x in {3.2,4.7,...,15.3} {\filldraw[draw=fzinkii, line width=0.4pt, fill=fzbump] (\x,9.45) circle[radius=0.42];}
  \foreach \x in {20.0,21.5,...,30.6} {\filldraw[draw=fzinkii, line width=0.4pt, fill=fzbump] (\x,9.45) circle[radius=0.42];}
  \filldraw[draw=fzinkii, line width=0.4pt, fill=fzdie] (2.2,9.9) rectangle (16.2,15.2);
  \filldraw[draw=fzinkii, line width=0.4pt, fill=fzdie] (19.2,9.9) rectangle (32.2,11.3);
  \foreach \y in {11.55,12.85,14.15,15.45} {\filldraw[draw=fzinkii, line width=0.4pt, fill=fzdie] (19.2,\y) rectangle (32.2,\y+1.05);}
  \foreach \x in {22.4,25.7,29.0} {\draw[fztsv, line width=0.5pt] (\x,11.3) -- (\x,16.5);}
  \node[fznote] at (9.2,12.55) {chiplet};
  \node[fznote, anchor=south] at (25.7,16.55) {die stack};
  \node[fznote] at (7.4,6.8) {interposer};
  \node[fznote] at (6.2,1.5) {substrate};
  \draw[fzcudk, line width=1.5pt, line cap=round] (15.2,9.45) -- (15.2,8.6) -- (20.0,8.6) -- (20.0,9.45);
  \draw[fzcu, line width=0.75pt, line cap=round] (15.2,9.45) -- (15.2,8.6) -- (20.0,8.6) -- (20.0,9.45);
}

\newcommand{\fzPatchTraces}[2]{%
  \colorlet{fztmpB}{#2}%
  \foreach \tr in {{(3.2,8.4) -- (10.2,8.4) -- (12.2,6.4) -- (20.8,6.4)},{(3.2,5.4) -- (8.6,5.4) -- (10.6,3.4) -- (20.8,3.4)}} {
    \draw[#2, line width=2.9pt, line cap=round, line join=round] \tr;
    \draw[#1, line width=2.1pt, line cap=round, line join=round] \tr;}
  \foreach \p in {(3.2,8.4),(20.8,6.4),(3.2,5.4),(20.8,3.4)} {\filldraw[draw=#2, line width=0.4pt, fill=#1] \p circle[radius=1.0];}
  \filldraw[draw=#2, line width=0.4pt, fill=#1] (15.4,3.4) circle[radius=0.8];
  \filldraw[draw=#2, line width=0.4pt, fill=fztmpB!70] (15.4,3.4) circle[radius=0.36];}
\newcommand{\fzPatchGround}[1]{%
  \fill[#1] (0,10.6) rectangle (24,11.8);  \fill[#1] (0,0.2) rectangle (24,1.4);}
\newcommand{\fzPatchPorts}{%
  \foreach \p/\q in {{(0.7,8.4)}/{(2.2,8.4)},{(0.7,5.4)}/{(2.2,5.4)},{(23.3,6.4)}/{(21.8,6.4)},{(23.3,3.4)}/{(21.8,3.4)}} {
    \draw[fzink, line width=0.4pt] \p -- \q; \fill[fzink] \p ++(-0.55,-0.55) rectangle ++(1.1,1.1);}}

\newcommand{\fzGraph}{%
  \coordinate (q1) at (0.9,8.4);  \coordinate (q2) at (0.9,5.4);  \coordinate (q3) at (23.1,6.4); \coordinate (q4) at (23.1,3.4);
  \coordinate (d1) at (3.2,8.4);  \coordinate (d2) at (3.2,5.4);  \coordinate (d3) at (20.8,6.4); \coordinate (d4) at (20.8,3.4);
  \coordinate (a1) at (6.7,8.4);  \coordinate (a2) at (11.2,7.4); \coordinate (a3) at (16.0,6.4);
  \coordinate (b1) at (5.9,5.4);  \coordinate (b2) at (9.6,4.4);  \coordinate (b3) at (12.9,3.4); \coordinate (v)  at (15.4,3.4);
  \coordinate (b4) at (18.1,3.4);
  \coordinate (r1) at (6.3,11.2); \coordinate (r2) at (17.7,11.2); \coordinate (r3) at (6.3,0.8);  \coordinate (r4) at (17.7,0.8);
  \draw[fzeref] (a1) -- (r1) (a3) -- (r2) (b1) -- (r3) (b4) -- (r4) (b3) -- (r4);
  \draw[fzecpl] (a1) -- (b1) (a2) -- (b2) (a3) -- (b3) (a3) -- (b4);
  \draw[fzeser] (q1) -- (d1) -- (a1) -- (a2) -- (a3) -- (d3) -- (q3)  (q2) -- (d2) -- (b1) -- (b2) -- (b3) -- (v) -- (b4) -- (d4) -- (q4);
  \foreach \c in {r1,r2,r3,r4} {\node[fznref] at (\c) {};}
  \foreach \c in {a1,a2,a3,b1,b2,b3,b4} {\node[fzntrace] at (\c) {};}
  \foreach \c in {d1,d2,d3,d4} {\node[fznpad] at (\c) {};}
  \node[fznvia] at (v) {};
  \foreach \c in {q1,q2,q3,q4} {\node[fznport] at (\c) {};}
}

\pgfmathdeclarefunction{fzSa}{1}{\pgfmathparse{0.93 - 0.30*(#1+0.001)^1.4 + 0.035*sin(1250*#1)}}
\pgfmathdeclarefunction{fzSb}{1}{\pgfmathparse{0.07 + 0.17*#1 + 0.07*(sin(560*#1+15))^2}}
\pgfmathdeclarefunction{fzSraw}{1}{\pgfmathparse{0.925 - 0.10*#1 + 0.20*exp(-((#1-0.60)/0.075)^2)}}
\pgfmathdeclarefunction{fzSenf}{1}{\pgfmathparse{0.905 - 0.10*#1 + 0.085*exp(-((#1-0.60)/0.10)^2) - 0.014*sin(900*#1)}}
\pgfmathdeclarefunction{fzSours}{1}{\pgfmathparse{0.90 - 0.11*#1 + 0.05*exp(-((#1-0.60)/0.09)^2) + 0.01*sin(700*#1)}}
\newcommand{\fzAxes}[2]{%
  \draw[fzinkii, line width=0.4pt, line cap=round] (0,#2+0.5) -- (0,0) -- (#1+0.6,0);
  \node[fznote, anchor=south west, inner sep=0.4pt] at (#1+0.15,0.15) {\fzm{f}};}

\newcommand{\fzSigma}[3]{%
  \pgfmathsetmacro{\fzone}{#2*(1-0.70)/0.42}%
  \fzAxes{#1}{#2}
  \ifnum#3=1
    \begin{scope}
      \clip (0,\fzone) rectangle (#1,#2+1);
      \fill[fzvf!30] plot[domain=0:1, samples=90] ({#1*\x},{#2*(fzSraw(\x)-0.70)/0.42}) -- (#1,0) -- (0,0) -- cycle;
    \end{scope}
  \fi
  \draw[fzinkii, line width=0.4pt, dash pattern=on 1.4pt off 1pt] (0,\fzone) -- (#1,\fzone);
  \node[fznote, anchor=east, inner sep=0.8pt] at (0,\fzone) {\fzm{1}};
  \node[fznote, anchor=south west, inner sep=0.4pt] at (0.35,#2-0.35) {\fzm{\sigma_{\fzs{\max}}}};
  \ifnum#3=1
    \draw[fzvf, line width=0.75pt] plot[domain=0:1, samples=90] ({#1*\x},{#2*(fzSraw(\x)-0.70)/0.42});
    \draw[fzink, line width=0.7pt] plot[domain=0:1, samples=90] ({#1*\x},{#2*(fzSenf(\x)-0.70)/0.42});
    \draw[fzink, line width=0.45pt, -{Stealth[length=2.2pt,width=2.0pt]}] ({0.6*#1+1.9},{#2*(fzSraw(0.6)-0.70)/0.42+0.2}) -- ({0.6*#1+1.9},{#2*(fzSenf(0.6)-0.70)/0.42+0.45});
  \else
    \draw[fzoursdk, line width=0.75pt] plot[domain=0:1, samples=90] ({#1*\x},{#2*(fzSours(\x)-0.70)/0.42});
  \fi}

\newcommand{\fzBrk}[4]{%
  \draw[fzinkii, line width=0.5pt] ({#1+0.7},{#2+0.45}) -- ({#1-0.45},{#2+0.45}) -- ({#1-0.45},{#4-0.45}) -- ({#1+0.7},{#4-0.45});
  \draw[fzinkii, line width=0.5pt] ({#3-0.7},{#2+0.45}) -- ({#3+0.45},{#2+0.45}) -- ({#3+0.45},{#4-0.45}) -- ({#3-0.7},{#4-0.45});}

\tikzset{
  fzb16/.style={fill=fzkA}, fzb24/.style={fill=fzkB}, fzb32/.style={fill=fzkC},
  fzbflat/.style={fill=fzkB}, fzbpale/.style={fill=fzkC},
  fzbnew/.style={fill=white, postaction={pattern={Lines[angle=45,distance=1.45pt,line width=0.4pt]}, pattern color=fzoursdk!80}},
  fzmline/.style={draw=fzoursdk, line width=0.4pt},
  fzpline/.style={draw=fzinkii, line width=0.4pt},
  fzhair/.style={draw=fzref, line width=0.4pt},
  fzop/.style={rounded rectangle, draw=fzoursdk, line width=0.5pt, fill=white, font=\footnotesize,
               inner xsep=2pt, inner ysep=1.2pt, minimum height=4.2mm},
  fzpanel/.style={fill=fzoband, draw=fzours!40, line width=0.4pt, rounded corners=3pt},
  fzhead/.style={draw=fzoursdk, line width=0.45pt, fill=white, rounded corners=0.6pt, font=\footnotesize,
                 minimum width=5.0mm, minimum height=3.2mm, inner sep=0pt},
}
\newcommand{\fzBands}[8]{%
  \path[#6] (#1,#3) rectangle (#2,{#3-16*#4});
  \ifnum#5>16 \path[#7] (#1,{#3-16*#4}) rectangle (#2,{#3-24*#4}); \fi
  \ifnum#5>24 \path[#8] (#1,{#3-24*#4}) rectangle (#2,{#3-32*#4}); \fi}
\newif\iffzZero \fzZerotrue
\def\fzUfill{}%
\newcommand{\fzL}[6]{%
  \pgfmathsetmacro{\fzS}{#3*#1}%
  \fill[white] (0,0) rectangle (#2+\fzS,-#2-\fzS);
  \fill[fzport] (0,0) -- (0,-#2) -- (#2,-#2) -- cycle;
  \ifx\fzUfill\empty \fzBands{0}{#2}{-#2}{#1}{#3}{#4}{#5}{#6}\else \path[\fzUfill] (0,-#2) rectangle (#2,-#2-\fzS);\fi
  \begin{scope}\clip (#2,-#2) -- (#2,-#2-\fzS) -- (#2+\fzS,-#2-\fzS) -- cycle;
    \fzBands{#2}{#2+\fzS}{-#2}{#1}{#3}{#4}{#5}{#6}\end{scope}
  \draw[fzhair] (0,-#2) -- (#2+\fzS,-#2) (#2,0) -- (#2,-#2-\fzS);
  \draw[fzinkii, line width=0.4pt] (0,0) -- (#2,-#2);
  \draw[fzoursdk, line width=0.55pt] (#2,-#2) -- (#2+\fzS,-#2-\fzS);
  \draw[fzmline] (0,0) rectangle (#2+\fzS,-#2-\fzS);
  \iffzZero \node[fzmath, text=fzinkii] at (#2+\fzS-1.45,-#2/2) {$0$};\fi
  \fzBrk{0}{0}{#2+\fzS}{-#2-\fzS}}
\newcommand{\fzLframes}[3]{%
  \ifnum#3>16 \draw[fzoursdk, line width=0.6pt] (0,0) rectangle (#2+16*#1,-#2-16*#1); \fi
  \ifnum#3>24 \draw[fzoursdk, line width=0.55pt, dash pattern=on 1.3pt off 0.8pt] (0,0) rectangle (#2+24*#1,-#2-24*#1); \fi}

\newcommand{\fzBar}[6]{%
  \fzBands{0}{#3}{0}{#1}{#2}{#4}{#5}{#6}
  \draw[fzmline] (0,0) rectangle (#3,{-#2*#1});}

\pgfmathdeclarefunction{fzSig}{1}{\pgfmathparse{(#1<-0.2)?0:((#1>0.2)?1:(0.5+0.5*sin(450*#1)))}}

\def\fzMeshEdges{(0.00,0.00)--(0.00,0.20) (0.00,0.00)--(3.00,0.00) (0.00,0.00)--(3.00,0.20) (0.00,0.20)--(0.00,1.40) (0.00,0.20)--(3.00,0.20) (0.00,1.40)--(0.00,2.00) (0.00,1.40)--(3.00,0.20) (0.00,1.40)--(3.00,1.40) (0.00,2.00)--(0.00,4.00) (0.00,2.00)--(1.58,3.51) (0.00,2.00)--(3.00,1.40) (0.00,4.00)--(0.00,6.00) (0.00,4.00)--(1.58,3.51) (0.00,4.00)--(2.20,5.40) (0.00,6.00)--(0.00,8.00) (0.00,6.00)--(2.20,5.40) (0.00,6.00)--(2.70,6.27) (0.00,6.00)--(2.70,7.53) (0.00,8.00)--(0.00,10.00) (0.00,8.00)--(2.20,8.40) (0.00,8.00)--(2.70,7.53) (0.00,10.00)--(0.00,10.60) (0.00,10.00)--(2.20,8.40) (0.00,10.00)--(2.70,9.27) (0.00,10.00)--(3.00,10.60) (0.00,10.60)--(0.00,11.80) (0.00,10.60)--(3.00,10.60) (0.00,11.80)--(0.00,12.00) (0.00,11.80)--(3.00,10.60) (0.00,11.80)--(3.00,11.80) (0.00,11.80)--(3.00,12.00) (0.00,12.00)--(3.00,12.00) (1.58,3.51)--(2.20,5.40) (1.58,3.51)--(2.70,4.53) (1.58,3.51)--(3.00,1.40) (2.20,5.40)--(2.70,4.53) (2.20,5.40)--(2.70,6.27) (2.20,5.40)--(3.20,4.95) (2.20,5.40)--(3.20,5.85) (2.20,8.40)--(2.70,7.53) (2.20,8.40)--(2.70,9.27) (2.20,8.40)--(3.20,7.95) (2.20,8.40)--(3.20,8.85) (2.70,4.53)--(3.00,1.40) (2.70,4.53)--(3.20,4.95) (2.70,4.53)--(3.70,4.53) (2.70,6.27)--(2.70,7.53) (2.70,6.27)--(3.20,5.85) (2.70,6.27)--(3.70,6.27) (2.70,6.27)--(3.70,7.53) (2.70,7.53)--(3.20,7.95) (2.70,7.53)--(3.70,7.53) (2.70,9.27)--(3.00,10.60) (2.70,9.27)--(3.20,8.85) (2.70,9.27)--(3.70,9.27) (3.00,0.00)--(3.00,0.20) (3.00,0.00)--(6.00,0.00) (3.00,0.20)--(3.00,1.40) (3.00,0.20)--(6.00,0.00) (3.00,0.20)--(6.00,0.20) (3.00,0.20)--(6.00,1.40) (3.00,1.40)--(3.70,4.53) (3.00,1.40)--(5.67,3.62) (3.00,1.40)--(6.00,1.40) (3.00,10.60)--(3.00,11.80) (3.00,10.60)--(3.70,9.27) (3.00,10.60)--(6.00,10.60) (3.00,11.80)--(3.00,12.00) (3.00,11.80)--(6.00,10.60) (3.00,11.80)--(6.00,11.80) (3.00,11.80)--(6.00,12.00) (3.00,12.00)--(6.00,12.00) (3.20,4.95)--(3.20,5.85) (3.20,4.95)--(3.70,4.53) (3.20,4.95)--(4.20,5.40) (3.20,5.85)--(3.70,6.27) (3.20,5.85)--(4.20,5.40) (3.20,7.95)--(3.20,8.85) (3.20,7.95)--(3.70,7.53) (3.20,7.95)--(4.20,8.40) (3.20,8.85)--(3.70,9.27) (3.20,8.85)--(4.20,8.40) (3.70,4.53)--(4.20,5.40) (3.70,4.53)--(5.00,4.95) (3.70,4.53)--(5.67,3.62) (3.70,6.27)--(3.70,7.53) (3.70,6.27)--(4.20,5.40) (3.70,6.27)--(4.95,7.95) (3.70,6.27)--(5.00,5.85) (3.70,7.53)--(4.20,8.40) (3.70,7.53)--(4.95,7.95) (3.70,9.27)--(4.20,8.40) (3.70,9.27)--(4.95,8.85) (3.70,9.27)--(6.00,10.60) (4.20,5.40)--(5.00,4.95) (4.20,5.40)--(5.00,5.85) (4.20,8.40)--(4.95,7.95) (4.20,8.40)--(4.95,8.85) (4.95,7.95)--(4.95,8.85) (4.95,7.95)--(5.00,5.85) (4.95,7.95)--(6.70,7.95) (4.95,7.95)--(6.70,8.85) (4.95,8.85)--(6.00,10.60) (4.95,8.85)--(6.70,8.85) (5.00,4.95)--(5.00,5.85) (5.00,4.95)--(5.67,3.62) (5.00,4.95)--(6.80,4.95) (5.00,4.95)--(6.80,5.85) (5.00,5.85)--(6.70,7.95) (5.00,5.85)--(6.80,5.85) (5.67,3.62)--(6.00,1.40) (5.67,3.62)--(6.80,4.95) (5.67,3.62)--(7.56,3.30) (6.00,0.00)--(6.00,0.20) (6.00,0.00)--(9.00,0.00) (6.00,0.20)--(6.00,1.40) (6.00,0.20)--(9.00,0.00) (6.00,0.20)--(9.00,0.20) (6.00,1.40)--(7.56,3.30) (6.00,1.40)--(9.00,0.20) (6.00,1.40)--(9.00,1.40) (6.00,10.60)--(6.00,11.80) (6.00,10.60)--(6.70,8.85) (6.00,10.60)--(9.00,10.60) (6.00,10.60)--(9.00,11.80) (6.00,11.80)--(6.00,12.00) (6.00,11.80)--(9.00,11.80) (6.00,12.00)--(9.00,11.80) (6.00,12.00)--(9.00,12.00) (6.70,7.95)--(6.70,8.85) (6.70,7.95)--(6.80,5.85) (6.70,7.95)--(8.45,7.95) (6.70,8.85)--(8.45,7.95) (6.70,8.85)--(8.45,8.85) (6.70,8.85)--(9.00,10.60) (6.80,4.95)--(6.80,5.85) (6.80,4.95)--(7.56,3.30) (6.80,4.95)--(8.28,5.08) (6.80,5.85)--(8.28,5.08) (6.80,5.85)--(8.45,7.95) (6.80,5.85)--(8.60,5.85) (7.56,3.30)--(8.28,5.08) (7.56,3.30)--(8.60,4.95) (7.56,3.30)--(9.00,1.40) (7.56,3.30)--(10.28,3.08) (8.28,5.08)--(8.60,4.95) (8.28,5.08)--(8.60,5.85) (8.45,7.95)--(8.45,8.85) (8.45,7.95)--(8.60,5.85) (8.45,7.95)--(8.92,5.72) (8.45,7.95)--(9.88,8.08) (8.45,8.85)--(9.00,10.60) (8.45,8.85)--(9.88,8.08) (8.45,8.85)--(10.20,8.85) (8.60,4.95)--(8.60,5.85) (8.60,4.95)--(8.92,5.72) (8.60,4.95)--(10.28,3.08) (8.60,4.95)--(10.60,3.85) (8.60,5.85)--(8.92,5.72) (8.92,5.72)--(9.88,8.08) (8.92,5.72)--(10.20,7.95) (8.92,5.72)--(10.60,3.85) (8.92,5.72)--(11.88,6.08) (9.00,0.00)--(9.00,0.20) (9.00,0.00)--(12.00,0.00) (9.00,0.00)--(12.00,0.20) (9.00,0.20)--(9.00,1.40) (9.00,0.20)--(12.00,0.20) (9.00,1.40)--(10.28,3.08) (9.00,1.40)--(10.60,2.95) (9.00,1.40)--(12.00,0.20) (9.00,1.40)--(12.00,1.40) (9.00,10.60)--(9.00,11.80) (9.00,10.60)--(10.20,8.85) (9.00,10.60)--(12.00,10.60) (9.00,11.80)--(9.00,12.00) (9.00,11.80)--(12.00,10.60) (9.00,11.80)--(12.00,11.80) (9.00,11.80)--(12.00,12.00) (9.00,12.00)--(12.00,12.00) (9.88,8.08)--(10.20,7.95) (9.88,8.08)--(10.20,8.85) (10.20,7.95)--(10.20,8.85) (10.20,7.95)--(10.52,8.72) (10.20,7.95)--(11.88,6.08) (10.20,7.95)--(12.20,6.85) (10.20,8.85)--(10.52,8.72) (10.20,8.85)--(12.00,10.60) (10.28,3.08)--(10.60,2.95) (10.28,3.08)--(10.60,3.85) (10.52,8.72)--(12.00,10.60) (10.52,8.72)--(12.20,6.85) (10.52,8.72)--(12.70,8.73) (10.60,2.95)--(10.60,3.85) (10.60,2.95)--(10.92,3.72) (10.60,2.95)--(12.00,1.40) (10.60,2.95)--(12.64,2.95) (10.60,3.85)--(10.92,3.72) (10.60,3.85)--(11.88,6.08) (10.92,3.72)--(11.88,6.08) (10.92,3.72)--(12.20,5.95) (10.92,3.72)--(12.64,2.95) (10.92,3.72)--(12.64,3.85) (11.88,6.08)--(12.20,5.95) (11.88,6.08)--(12.20,6.85) (12.00,0.00)--(12.00,0.20) (12.00,0.00)--(15.00,0.00) (12.00,0.20)--(12.00,1.40) (12.00,0.20)--(15.00,0.00) (12.00,0.20)--(15.00,0.20) (12.00,0.20)--(15.00,1.40) (12.00,1.40)--(12.64,2.95) (12.00,1.40)--(15.00,1.40) (12.00,10.60)--(12.00,11.80) (12.00,10.60)--(12.70,8.73) (12.00,10.60)--(15.00,10.60) (12.00,11.80)--(12.00,12.00) (12.00,11.80)--(15.00,10.60) (12.00,11.80)--(15.00,11.80) (12.00,11.80)--(15.00,12.00) (12.00,12.00)--(15.00,12.00) (12.20,5.95)--(12.20,6.85) (12.20,5.95)--(12.52,6.72) (12.20,5.95)--(12.64,3.85) (12.20,5.95)--(13.92,5.95) (12.20,6.85)--(12.52,6.72) (12.20,6.85)--(12.70,8.73) (12.52,6.72)--(12.70,8.73) (12.52,6.72)--(13.92,5.95) (12.52,6.72)--(13.92,6.85) (12.64,2.95)--(12.64,3.85) (12.64,2.95)--(14.60,3.40) (12.64,2.95)--(14.68,2.95) (12.64,2.95)--(15.00,1.40) (12.64,3.85)--(13.92,5.95) (12.64,3.85)--(14.60,3.40) (12.64,3.85)--(14.68,3.85) (12.70,8.73)--(13.92,6.85) (12.70,8.73)--(14.92,8.83) (12.70,8.73)--(15.00,10.60) (13.92,5.95)--(13.92,6.85) (13.92,5.95)--(14.68,3.85) (13.92,5.95)--(15.00,4.09) (13.92,5.95)--(15.64,5.95) (13.92,6.85)--(14.92,8.83) (13.92,6.85)--(15.64,5.95) (13.92,6.85)--(15.64,6.85) (14.60,3.40)--(14.68,2.95) (14.60,3.40)--(14.68,3.85) (14.60,3.40)--(15.00,2.71) (14.60,3.40)--(15.00,4.09) (14.68,2.95)--(15.00,1.40) (14.68,2.95)--(15.00,2.71) (14.68,3.85)--(15.00,4.09) (14.92,8.83)--(15.00,10.60) (14.92,8.83)--(15.64,6.85) (14.92,8.83)--(17.06,9.33) (15.00,0.00)--(15.00,0.20) (15.00,0.00)--(18.00,0.00) (15.00,0.20)--(15.00,1.40) (15.00,0.20)--(18.00,0.00) (15.00,0.20)--(18.00,0.20) (15.00,1.40)--(15.00,2.71) (15.00,1.40)--(15.80,2.71) (15.00,1.40)--(18.00,0.20) (15.00,1.40)--(18.00,1.40) (15.00,2.71)--(15.00,4.09) (15.00,2.71)--(15.80,2.71) (15.00,2.71)--(15.80,4.09) (15.00,4.09)--(15.64,5.95) (15.00,4.09)--(15.80,4.09) (15.00,10.60)--(15.00,11.80) (15.00,10.60)--(17.06,9.33) (15.00,10.60)--(18.00,10.60) (15.00,10.60)--(18.00,11.80) (15.00,11.80)--(15.00,12.00) (15.00,11.80)--(18.00,11.80) (15.00,12.00)--(18.00,11.80) (15.00,12.00)--(18.00,12.00) (15.64,5.95)--(15.64,6.85) (15.64,5.95)--(15.80,4.09) (15.64,5.95)--(17.36,5.95) (15.64,5.95)--(17.36,6.85) (15.64,6.85)--(17.06,9.33) (15.64,6.85)--(17.36,6.85) (15.80,2.71)--(15.80,4.09) (15.80,2.71)--(16.20,3.40) (15.80,2.71)--(16.72,2.95) (15.80,2.71)--(18.00,1.40) (15.80,4.09)--(16.20,3.40) (15.80,4.09)--(16.72,3.85) (15.80,4.09)--(17.36,5.95) (16.20,3.40)--(16.72,2.95) (16.20,3.40)--(16.72,3.85) (16.72,2.95)--(16.72,3.85) (16.72,2.95)--(18.00,1.40) (16.72,2.95)--(18.76,2.95) (16.72,3.85)--(17.36,5.95) (16.72,3.85)--(18.76,2.95) (16.72,3.85)--(18.76,3.85) (17.06,9.33)--(17.36,6.85) (17.06,9.33)--(18.00,10.60) (17.06,9.33)--(19.42,9.07) (17.36,5.95)--(17.36,6.85) (17.36,5.95)--(18.76,3.85) (17.36,5.95)--(19.08,5.95) (17.36,6.85)--(19.08,5.95) (17.36,6.85)--(19.08,6.85) (17.36,6.85)--(19.42,9.07) (18.00,0.00)--(18.00,0.20) (18.00,0.00)--(21.00,0.00) (18.00,0.20)--(18.00,1.40) (18.00,0.20)--(21.00,0.00) (18.00,0.20)--(21.00,0.20) (18.00,0.20)--(21.00,1.40) (18.00,1.40)--(18.76,2.95) (18.00,1.40)--(20.30,2.53) (18.00,1.40)--(21.00,1.40) (18.00,10.60)--(18.00,11.80) (18.00,10.60)--(19.42,9.07) (18.00,10.60)--(21.00,10.60) (18.00,10.60)--(21.00,11.80) (18.00,11.80)--(18.00,12.00) (18.00,11.80)--(21.00,11.80) (18.00,11.80)--(21.00,12.00) (18.00,12.00)--(21.00,12.00) (18.76,2.95)--(18.76,3.85) (18.76,2.95)--(19.80,3.40) (18.76,2.95)--(20.30,2.53) (18.76,3.85)--(19.08,5.95) (18.76,3.85)--(19.80,3.40) (18.76,3.85)--(20.30,4.27) (19.08,5.95)--(19.08,6.85) (19.08,5.95)--(19.80,6.40) (19.08,5.95)--(20.30,4.27) (19.08,5.95)--(20.30,5.53) (19.08,6.85)--(19.42,9.07) (19.08,6.85)--(19.80,6.40) (19.08,6.85)--(20.30,7.27) (19.42,9.07)--(20.30,7.27) (19.42,9.07)--(21.00,10.60) (19.42,9.07)--(21.30,7.27) (19.80,3.40)--(20.30,2.53) (19.80,3.40)--(20.30,4.27) (19.80,3.40)--(20.80,2.95) (19.80,3.40)--(20.80,3.85) (19.80,6.40)--(20.30,5.53) (19.80,6.40)--(20.30,7.27) (19.80,6.40)--(20.80,5.95) (19.80,6.40)--(20.80,6.85) (20.30,2.53)--(20.80,2.95) (20.30,2.53)--(21.00,1.40) (20.30,2.53)--(21.30,2.53) (20.30,4.27)--(20.30,5.53) (20.30,4.27)--(20.80,3.85) (20.30,4.27)--(21.30,4.27) (20.30,5.53)--(20.80,5.95) (20.30,5.53)--(21.30,4.27) (20.30,5.53)--(21.30,5.53) (20.30,7.27)--(20.80,6.85) (20.30,7.27)--(21.30,7.27) (20.80,2.95)--(20.80,3.85) (20.80,2.95)--(21.30,2.53) (20.80,2.95)--(21.80,3.40) (20.80,3.85)--(21.30,4.27) (20.80,3.85)--(21.80,3.40) (20.80,5.95)--(20.80,6.85) (20.80,5.95)--(21.30,5.53) (20.80,5.95)--(21.80,6.40) (20.80,6.85)--(21.30,7.27) (20.80,6.85)--(21.80,6.40) (21.00,0.00)--(21.00,0.20) (21.00,0.00)--(24.00,0.00) (21.00,0.00)--(24.00,0.20) (21.00,0.20)--(21.00,1.40) (21.00,0.20)--(24.00,0.20) (21.00,1.40)--(21.30,2.53) (21.00,1.40)--(24.00,0.20) (21.00,1.40)--(24.00,1.40) (21.00,10.60)--(21.00,11.80) (21.00,10.60)--(21.30,7.27) (21.00,10.60)--(24.00,10.00) (21.00,10.60)--(24.00,10.60) (21.00,10.60)--(24.00,11.80) (21.00,11.80)--(21.00,12.00) (21.00,11.80)--(24.00,11.80) (21.00,12.00)--(24.00,11.80) (21.00,12.00)--(24.00,12.00) (21.30,2.53)--(21.80,3.40) (21.30,2.53)--(24.00,1.40) (21.30,2.53)--(24.00,2.00) (21.30,4.27)--(21.30,5.53) (21.30,4.27)--(21.80,3.40) (21.30,4.27)--(24.00,4.00) (21.30,5.53)--(21.80,6.40) (21.30,5.53)--(24.00,4.00) (21.30,5.53)--(24.00,6.00) (21.30,7.27)--(21.80,6.40) (21.30,7.27)--(24.00,8.00) (21.30,7.27)--(24.00,10.00) (21.80,3.40)--(24.00,2.00) (21.80,3.40)--(24.00,4.00) (21.80,6.40)--(24.00,6.00) (21.80,6.40)--(24.00,8.00) (24.00,0.00)--(24.00,0.20) (24.00,0.20)--(24.00,1.40) (24.00,1.40)--(24.00,2.00) (24.00,2.00)--(24.00,4.00) (24.00,4.00)--(24.00,6.00) (24.00,6.00)--(24.00,8.00) (24.00,8.00)--(24.00,10.00) (24.00,10.00)--(24.00,10.60) (24.00,10.60)--(24.00,11.80) (24.00,11.80)--(24.00,12.00)}

\tikzset{
  fzcut/.style={draw=fzoursdk, line width=0.7pt, dash pattern=on 2pt off 1pt},
  fznums/.style={fznum, minimum size=2.45mm, line width=0.45pt},
  fzfbox/.style={rounded corners=1.2pt, draw=fzoursdk, line width=0.5pt, fill=white, font=\footnotesize,
                 inner xsep=1.4pt, inner ysep=1.1pt, minimum height=4.6mm},
}
\newcommand{\fzstep}[1]{\tikz[baseline=(fzsn.base)]\node[fznum](fzsn){#1};\,}

\tikzset{
  fzpart/.style={draw=fzref, line width=0.4pt},                    %
  fzbtag/.style={font=\scriptsize, text=fzinkii, inner sep=0.3pt},  %
  fzbus/.style={draw=fzinkii, line width=0.6pt, line cap=round, line join=round},
  fzres/.style={draw=fzinkii, line width=0.5pt, line cap=butt},     %
}
\renewcommand{\fzstep}[1]{\tikz[baseline=(fzsn.base)]\node[fznum](fzsn){#1};\hspace{0.9pt}\,}

\newcommand{\fzWone}[2]{%
  \fill[fzkB] (0,0) rectangle (#1+#2,-#1-#2);
  \fill[fzport] (0,0) rectangle (#1,-#1);
  \draw[fzpart] (#1,0) -- (#1,-#1-#2) (0,-#1) -- (#1+#2,-#1);
  \draw[fzmline, line width=0.5pt] (0,0) rectangle (#1+#2,-#1-#2);
  \node[fzmath] at (#1/2,-#1/2) {$D_{\fzs{0}}$};
  \node[fzmath] at (#1+#2/2,-#1/2) {$P\fzT$};
  \node[fzmath] at (#1/2,-#1-#2/2) {$P$};
  \node[fzmath] at (#1+#2/2,-#1-#2/2) {$R$};
  \fzBrk{0}{0}{#1+#2}{-#1-#2}}

\pgfmathdeclarefunction{fzSl}{1}{\pgfmathparse{0.78 - 0.30*(#1+0.001)^1.25 + 0.075*sin(620*#1+10)}}

\newcommand{\fzLamone}[6]{%
  \pgfmathsetmacro{\fzS}{#2*#1}%
  \fill[white] (0,0) rectangle (\fzS,-\fzS);
  \begin{scope}\clip (0,-#6) -- (0,-\fzS) -- (\fzS-#6,-\fzS) -- cycle;
    \fzBands{0}{\fzS}{0}{#1}{#2}{#3}{#4}{#5}\end{scope}
  \draw[fzref, line width=0.5pt, dash pattern=on 0.5pt off 0.75pt, line cap=round] (0.15,-0.15) -- (\fzS-0.15,-\fzS+0.15);
  \draw[fzmline] (0,-#6) -- (\fzS-#6,-\fzS);
  \draw[fzmline] (0,0) rectangle (\fzS,-\fzS);}
\newcommand{\fzABCDnest}[1]{%
  \pgfmathsetmacro{\fzU}{#1}\pgfmathsetmacro{\fzN}{32*\fzU}\pgfmathsetmacro{\fzM}{8*\fzU}\pgfmathsetmacro{\fzTot}{\fzN+\fzM}%
  \fill[fzkC] (0,0) rectangle (\fzTot,-\fzTot);
  \fill[fzkB] (0,0) rectangle (24*\fzU,-24*\fzU);  \fill[fzkA] (0,0) rectangle (16*\fzU,-16*\fzU);
  \fill[fzkA] (\fzN,0) rectangle (\fzTot,-16*\fzU);   \fill[fzkB] (\fzN,-16*\fzU) rectangle (\fzTot,-24*\fzU);
  \fill[fzkA] (0,-\fzN) rectangle (16*\fzU,-\fzTot);  \fill[fzkB] (16*\fzU,-\fzN) rectangle (24*\fzU,-\fzTot);
  \fill[fzport] (\fzN,-\fzN) rectangle (\fzTot,-\fzTot);
  \draw[fzpart] (\fzN,0) -- (\fzN,-\fzTot) (0,-\fzN) -- (\fzTot,-\fzN);
  \draw[fzoursdk, line width=0.5pt] (0,-16*\fzU) -| (16*\fzU,0)  (\fzN,-16*\fzU) -- (\fzTot,-16*\fzU)  (16*\fzU,-\fzN) -- (16*\fzU,-\fzTot);
  \draw[fzoursdk, line width=0.45pt, dash pattern=on 1.3pt off 0.8pt] (0,-24*\fzU) -| (24*\fzU,0)  (\fzN,-24*\fzU) -- (\fzTot,-24*\fzU)  (24*\fzU,-\fzN) -- (24*\fzU,-\fzTot);
  \draw[fzmline, line width=0.5pt] (0,0) rectangle (\fzTot,-\fzTot);
  \fzBrk{0}{0}{\fzTot}{-\fzTot}}

\pgfmathdeclarefunction{fzSone}{1}{\pgfmathparse{0.70 - 0.26*(#1+0.001)^1.4 + 0.04*sin(1100*#1+20)}}
\newcommand{\fzSplotOne}[2]{%
  \fzAxes{#1}{#2}
  \node[fznote, anchor=south west, inner sep=0.4pt] at (0.35,#2-0.35) {\fzm{|S_{\fzs{21}}|}};
  \draw[fzvf, line width=0.75pt] plot[domain=0:1, samples=80] ({#1*\x},{#2*fzSone(\x)});
  \foreach \t/\d in {0.05/-0.035, 0.17/0.035, 0.29/-0.035, 0.41/0.035, 0.53/-0.035, 0.65/0.035, 0.77/-0.035, 0.89/0.035} {
    \fill[fzinkii] ({#1*\t},{#2*(fzSone(\t)+\d)}) circle[radius=0.36];}
  \node[fznote, text=fzvfdk, anchor=base east, inner sep=0.4pt] at (#1,{#2*fzSone(1)+0.75}) {fit};
  \node[fznote, anchor=base, inner sep=0.4pt] at ({#1*0.33},1.0) {samples};}

\newcommand{\fzNetW}{%
  \foreach \k in {2,1,0} {\filldraw[draw=fzoursdk, line width=0.45pt, fill=white, rounded corners=0.8pt]
      (\k*0.8,-3.6+\k*0.8) rectangle (\k*0.8+5.0,3.0+\k*0.8);}
  \begin{scope}[shift={(2.5,-0.3)}]
    \draw[fzinkii, line width=0.4pt] (-1.4,-1.0) -- (0,1.3) -- (1.4,-0.6) -- cycle (0,1.3) -- (0.1,-0.1);
    \foreach \p in {(-1.4,-1.0),(0,1.3),(1.4,-0.6),(0.1,-0.1)} {\filldraw[draw=fzoursdk, fill=fzkB, line width=0.4pt] \p circle[radius=0.42];}
  \end{scope}
  \filldraw[draw=fzoursdk, line width=0.45pt, fill=fzkC, line join=round] (8.4,3.0) -- (11.0,1.6) -- (11.0,-2.2) -- (8.4,-3.6) -- cycle;
  \foreach \i in {0,...,4} {\filldraw[draw=fzoursdk, line width=0.4pt, fill=white] (12.0,1.6-\i*0.76) rectangle ++(1.2,-0.76);}
  \draw[fzinkii, line width=0.4pt, -{Stealth[length=2.2pt,width=2.0pt]}] (11.1,-0.3) -- (11.95,-0.3);
  \draw[fzinkii, line width=0.4pt, -{Stealth[length=2.2pt,width=2.0pt]}] (13.25,-0.3) -- (14.55,-0.3);
  \begin{scope}[shift={(15.3,2.25)}]\fzZerofalse\fzL{0.13}{1.3}{32}{fzbflat}{fzbflat}{fzbflat}\fzZerotrue\end{scope}
}
\usepackage{cite}
\usepackage{url}
\usepackage[hidelinks]{hyperref}
\usepackage{tabularx}
\hypersetup{pdfauthor={},pdftitle={Learning in the Passive Space: End-to-End Muti-Order Port-Hamiltonian Macromodeling}}

\newtheorem{proposition}{Proposition}

\theoremstyle{definition}
\newtheorem{problem}{Problem}

\begin{document}

\title{Learning in the Passive Space: End-to-End Muti-Order Port-Hamiltonian Macromodeling}

\author{
    \IEEEauthorblockN{Benchao Zhu, Keren Zhu\IEEEauthorrefmark{1}}
    \IEEEauthorblockA{College of Integrated Circuits and Micro-Nano Electronics \\ Fudan University, Shanghai, China \\
    zhubc25@m.fudan.edu.cn, krzhu@fudan.edu.cn}
}

\maketitle

\begin{abstract}
Accurate transient signal- and power-integrity simulation requires compact state-space macromodels that remain strictly passive across all frequencies. Conventional electromagnetic (EM) macromodeling requires minutes to hours of full-wave simulation, whereas both rational fitters and learned frequency-response surrogates rely on post-hoc fitting or passivity perturbation that degrades accuracy without guaranteeing full-spectrum passivity. We present a graph-to-state-space learning framework that optimizes directly within the passive space to map physical layouts to multi-order port-Hamiltonian (pH) macromodels. The network combines a heterogeneous layout-graph encoder with a complete, prefix-nested pH parameterization that assembles unconstrained outputs into a Lyapunov-stable, continuous-frequency passive realization, and slicing its leading $k$ states directly yields valid order-$k$ macromodels in one forward pass without post-hoc model order reduction. Using a geometric-mean multi-order loss and a progressive order-growth curriculum, we train all orders directly on observable S-parameters without reference state-space realizations. Experiments on 27k unseen test structures from a 270k-structure advanced-packaging dataset show that our method reduces mean S-parameter error by $42\%$ on average (up to $65\%$) compared with MFT-NNLS, MATLAB RF Toolbox, and Ansys Genequiv fitted to ground-truth EM responses, with zero passivity violations among all $459{,}000$ exported realizations. Each layout is mapped to macromodels in $36$--$85$~ms ($11{,}000$--$39{,}000\times$ faster than EM-based flows), and re-evaluation on denoised multi-solve EM references confirms reduced sensitivity to single-solve numerical jitter that per-structure fitters partially absorb.
\end{abstract}

\begin{figure*}[t]
\centering
\input{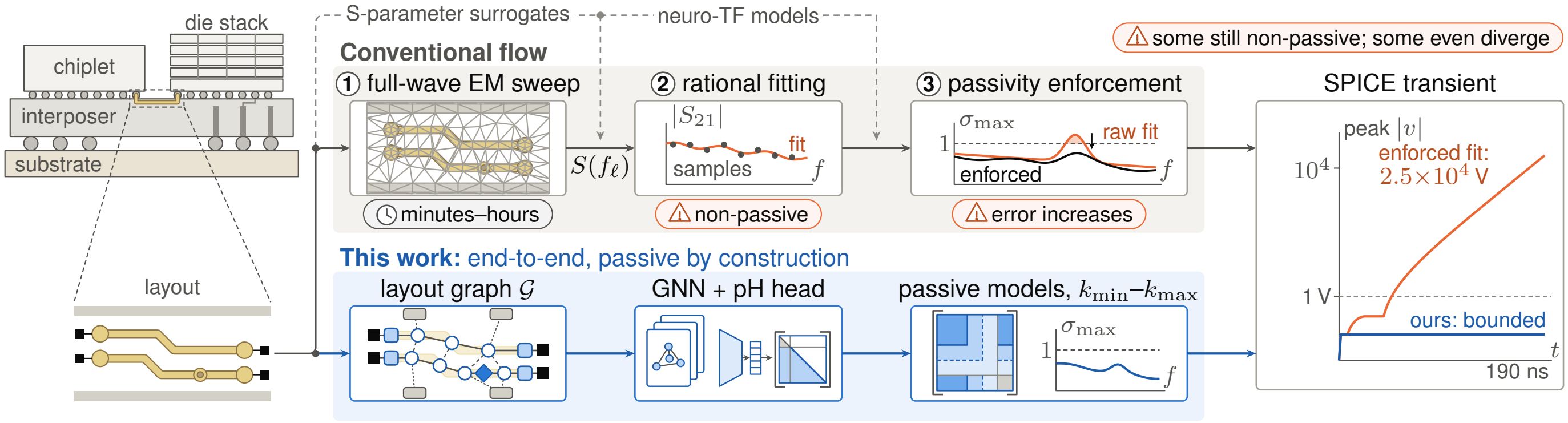}
\caption{Comparison of conventional macromodeling pipelines and the proposed end-to-end macromodeling framework. Top: Conventional modeling performs (1) a full-wave EM sweep, (2) rational fitting, and (3) perturbation-based passivity enforcement bounding the maximum singular value $\sigma_{\max}$. Dashed paths denote prior learning surrogates that bypass the sweep~\cite{traceformer,pcbformer,liu2024gnnsp,zhang2022pdncnn} or both the sweep and fitting~\cite{feng2016neurotf} yet still require passivity enforcement. Bottom: Trained on EM-simulated structures, our network maps an unseen layout graph directly to nested passive, stable realizations of $k_{\min}$ to $k_{\max}$ states in one pass. Right: Transient ngspice step response (1~V step via $50~\Omega$ into port~1, ports~2--8 open), where an enforced 32-state fit remains non-passive and diverges to $2.5\times 10^{4}\text{ V}$ within $190\text{ ns}$ while our model remains bounded.}
\label{figure:pipeline}
\end{figure*}

\section{Introduction}
\label{section:introduction}

Compact state-space macromodels are widely used to approximate the port behavior of interconnects and power-delivery networks (PDNs) in transient signal- and power-integrity (SI/PI) simulations~\cite{grivet2016book}. Reliable time-domain analysis requires these models to achieve high accuracy with few states~\cite{phillips2003prbt} while remaining strictly passive, as even minor violations can trigger simulation divergence under passive terminations~\cite{triverio2007stability}.

Conventional macromodeling follows the three-stage pipeline in Fig.~\ref{figure:pipeline} (top). First, a full-wave electromagnetic (EM) solver such as HFSS~\cite{hfss} computes scattering parameters (S-parameters) at discrete frequencies. Next, a rational model of a prescribed order is fitted to these frequency samples (e.g., via vector fitting~\cite{gustavsen1999vf}). Finally, passivity enforcement perturbs the fitted model until passivity criteria are satisfied within a given tolerance~\cite{gustavsen2008passivity,grivet2004hamiltonian}.

However, two bottlenecks limit the efficiency and reliability of this pipeline. First, computing a full-wave EM solution for each structure takes minutes to hours and must be repeated after every layout modification. Second, passivity is not intrinsic to the fitted model, but imposed via post-hoc perturbation, which degrades accuracy and may fail to converge. As shown in Fig.~\ref{figure:pipeline} (right), models can remain non-passive after passivity enforcement, leading to unstable transient simulations even with undriven ports left open.

Prior efforts typically mitigate one of these bottlenecks while leaving the other unresolved. Non-learning passive modeling techniques ensure passivity across frequencies, yet remain constrained by EM simulation cost or parameter dimensionality. While model order reduction preserves passivity when an RLC netlist is already available, passive fitting and refinement methods for sampled responses, such as convex fitting with fixed poles~\cite{coelho2004}, port-Hamiltonian (pH) fitting~\cite{benner2020phid}, and spectral-factor refinement~\cite{gspnn}, still require a full-wave EM simulation for every new layout. Parameterized macromodels~\cite{ferranti2009param,bradde2022uniform} and parametric pH reduction~\cite{schwerdtner2025psobmor} construct passive models over a design space, but rely on fixed basis functions or interpolation schemes that scale to only a few parameters.

On the other hand, learning-based approaches bypass EM simulations on unseen layouts, but fail to guarantee continuous-frequency passivity. Learned surrogates predict S- or Z-parameters directly from layouts~\cite{traceformer,pcbformer,liu2024gnnsp,zhang2022pdncnn}; however, their discrete frequency outputs still require rational fitting and passivity enforcement, even when output layers bound singular values~\cite{swaminathan_passive_nn}. Neuro-transfer-function (neuro-TF) models further eliminate post-fitting by predicting rational transfer functions from design variables~\cite{feng2016neurotf}, yet even with stable poles and per-sample singular-value scaling~\cite{xu2024deepotf}, passivity holds at most at the sampled frequencies rather than across the continuous spectrum.

Resolving these bottlenecks requires constraining learning itself to the passive space. Toward this goal, we propose an end-to-end macromodeling framework (Fig.~\ref{figure:pipeline}, bottom) that represents each layout as a graph and employs a graph neural network (GNN) with multilayer perceptrons (MLPs) to directly assemble a pH realization~\cite{vanderschaft2014ph} without test-time EM simulation, rational fitting, or post-hoc enforcement. Moreover, to serve structures of varying order requirements without separate per-order models or post-hoc reduction, each forward pass outputs nested realizations across multiple state counts. To the best of our knowledge, this is the first method to learn nested, continuous-frequency passive macromodels directly from layouts varying in technology, routing topology, return paths, and tens of geometric and material parameters. Our main contributions are as follows:

\begin{itemize}
\item \textbf{Complete passive parameterization.} We map unconstrained network outputs differentiably onto the full set of admissible pH realizations, enabling optimization directly within the passive space without loss penalties or post-hoc perturbations.
\item \textbf{Nested multi-order realization.} We fix the energy matrix to the identity and jointly factor the dissipation matrix so every leading $k$-state slice remains passive and invariant to appended states, yielding macromodels across orders $k \in \{k_{\min},\dots,k_{\max}\}$ in one forward pass.
\item \textbf{S-parameter order-growth curriculum.} We train directly on S-parameters without reference state-space fitting, pairing progressive state growth with a geometric-mean multi-order loss that prevents low-order residuals from dominating high-order refinement.
\end{itemize}

We evaluate the proposed framework on a dataset of 270k EM-simulated advanced-packaging structures spanning signal lanes, vertical transitions, and power-delivery meshes. Experimental results demonstrate that our networks generate passive macromodels from unseen layouts in $36$--$85$~ms ($11{,}000$--$39{,}000\times$ faster than EM-based flows), achieve zero passivity violations across all $459{,}000$ exported test realizations, and reduce mean S-parameter error by $42\%$ on average (up to $65\%$) over MFT-NNLS, MATLAB RF Toolbox, and Ansys Genequiv fitted to ground-truth EM responses. Evaluation against denoised multi-solve EM references further indicates that cross-layout learning mitigates sensitivity to single-solve numerical fluctuations that per-structure fitters partially absorb.

\section{Preliminaries}
\label{section:background}

In this section, we review state-space macromodels, introduce port-Hamiltonian systems, and formally state the problem.

\subsection{State-Space Macromodels}
\label{section:vf}

Consider a linear time-invariant (LTI) $m$-port structure represented in impedance form by an $n$-state realization:
\begin{equation}
\dot x=Ax+Bu,\quad y=Cx+Du,\quad Z(s)=C(sI-A)^{-1}B+D,
\label{equation:ss}
\end{equation}
where $x\in\mathbb{R}^n$ is the state vector, the input $u\in\mathbb{R}^m$ and the output $y\in\mathbb{R}^m$ collect the port currents and voltages, respectively, and $u^{\mathsf{T}} y$ represents the power entering the ports. For a reference impedance $Z_0>0$, the scattering matrix is given by the Cayley transform:
\begin{equation}
S(s)=\big(Z(s)-Z_0I\big)\big(Z(s)+Z_0I\big)^{-1}.
\label{equation:cayley}
\end{equation}

For zero input ($u = 0$), the realization~\eqref{equation:ss} is Lyapunov stable if every state trajectory remains bounded for all $t \ge 0$ (i.e., all eigenvalues of $A$ have nonpositive real parts, with any imaginary-axis eigenvalues being semisimple), and asymptotically (Hurwitz) stable if all eigenvalues of $A$ lie strictly in the open left half-plane. From zero initial state ($x(0) = 0$), the system is passive if the net absorbed energy satisfies $\int_0^t u(\tau)^{\mathsf{T}} y(\tau)\,d\tau \ge 0$ for all inputs $u$ and all $t \ge 0$. For real realizations, passivity is equivalent in the frequency domain to $Z(s)$ being positive real, or $S(s)$ being bounded real, i.e., analytic in $\operatorname{Re}(s)>0$ with $\|S(s)\|_2 \le 1$ (implying $\sigma_{\max}(S(j\omega)) = \|S(j\omega)\|_2 \le 1$ for all $\omega\in\mathbb{R}$). Passivity of a minimal realization implies Lyapunov stability, and further implies Hurwitz stability whenever $Z(s)$ has no purely imaginary (lossless) poles. The model order is defined as the state dimension $n$, which for a minimal realization equals the McMillan degree of $Z(s)$~\cite{kailath1980}.

\subsection{Port-Hamiltonian Systems}
\label{section:ph}

A linear port-Hamiltonian (pH) system~\cite{vanderschaft2014ph} is an $n$-state realization of the form
\begin{equation}
\dot x=(J-R)Qx+(G-P)u,\quad y=(G+P)^{\mathsf{T}} Qx+(D_0+N)u,
\label{equation:ph}
\end{equation}
whose parameter matrices satisfy
\begin{equation}
J=-J^{\mathsf{T}},\quad N=-N^{\mathsf{T}},\quad Q=Q^{\mathsf{T}}\succ 0,\quad
W=\begin{bmatrix}D_0 & P^{\mathsf{T}} \\ P & R\end{bmatrix}\succeq 0.
\label{equation:ph_cond}
\end{equation}
Here, $Q\in\mathbb{R}^{n\times n}$ is the energy matrix, $J\in\mathbb{R}^{n\times n}$ is the interconnection matrix, and $W\in\mathbb{R}^{(m+n)\times(m+n)}$ is the dissipation matrix comprising the symmetric port feedthrough $D_0\in\mathbb{R}^{m\times m}$, state dissipation $R\in\mathbb{R}^{n\times n}$, and port--state coupling $P\in\mathbb{R}^{n\times m}$. The skew-symmetric matrix $N\in\mathbb{R}^{m\times m}$ accounts for lossless feedthrough, while $G\in\mathbb{R}^{n\times m}$ is unconstrained. With the quadratic Hamiltonian $\mathcal{H}(x)=\frac{1}{2}x^{\mathsf{T}} Qx$, the skew-symmetry of $J$ and $N$ yields the power-balance relation
\begin{equation}
u^{\mathsf{T}} y - \dot{\mathcal{H}}(x) = \begin{bmatrix}u \\ Qx\end{bmatrix}^{\mathsf{T}} W \begin{bmatrix}u \\ Qx\end{bmatrix} \ge 0.
\end{equation}
Integrating from $x(0)=0$ gives $\int_0^t u(\tau)^{\mathsf{T}} y(\tau)\,d\tau \ge \mathcal{H}(x(t)) \ge 0$, ensuring passivity. Moreover, since $Q\succ 0$ and $R\succeq 0$, setting $u=0$ yields $\dot{\mathcal{H}}(x)=-x^{\mathsf{T}} QRQx \le 0$. Consequently, every realization satisfying~\eqref{equation:ph}--\eqref{equation:ph_cond}, whether minimal or not, is passive and Lyapunov stable, and is Hurwitz stable whenever no eigenvector of $JQ$ lies in $\ker(RQ)$ (in particular, whenever $R\succ 0$).

Conversely, every passive system with a minimal realization $(A,B,C,D)$ admits an equivalent pH realization of the same order $n$~\cite{beattie2019robust}. Specifically, one can retain the original state coordinates by selecting $Q\succ 0$ as a solution to the Kalman--Yakubovich--Popov (KYP) inequality~\cite{beattie2019robust}. Thus, parameterizing a macromodel via~\eqref{equation:ph}--\eqref{equation:ph_cond} guarantees passivity and stability by construction while preserving completeness over all minimal passive realizations.

\begin{figure*}[!htb]
\centering
\input{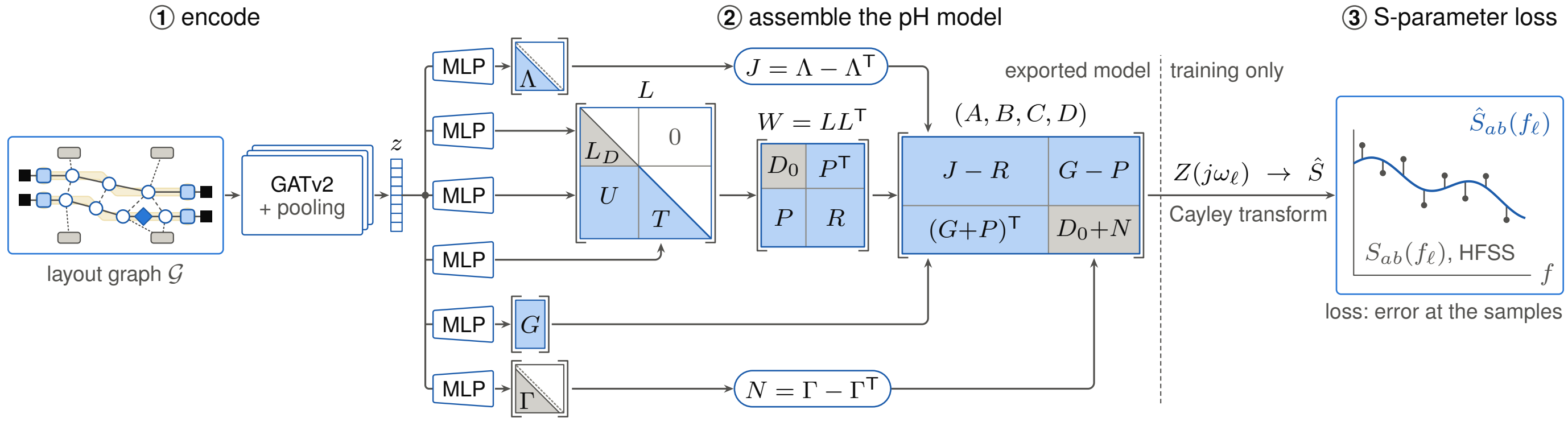}
\caption{Overview of the proposed multi-order passive macromodeling framework. (1)~A GATv2 encoder maps the layout graph $\mathcal{G}$ to a latent vector $z$. (2)~Parallel MLPs predict unconstrained blocks $(\Lambda,L_D,U,T,G,\Gamma)$, which deterministically assemble a $k_{\max}$-state pH realization $(A,B,C,D)$ with $Q=I$ (gray: order-shared port blocks; blue: state-indexed blocks). (3)~During training, evaluating $Z(j\omega_\ell)$ ($\omega_\ell=2\pi f_\ell$) and applying the Cayley transform yields $\hat{S}$ for direct S-parameter supervision against HFSS samples.}
\label{figure:overview}
\end{figure*}

\subsection{Problem Statement}
\label{section:problem}

Each physical structure in a design family is represented as a layout graph $\mathcal{G}$ encoding structural and physical attributes such as layout geometry, routing topology, return paths, and material and stack-up parameters. We formulate the multi-order macromodeling problem as follows.

\begin{problem}[Multi-Order Passive Macromodeling]
Given a family of $m$-port structures, an order range $\{k_{\min},\dots,k_{\max}\}$, and a training set of layout graphs paired with ground-truth frequency samples $\{S(f_\ell)\}_{\ell=1}^{N_f}$, learn a mapping that assigns to any layout graph $\mathcal{G}$ in the family a set of macromodels $\{\mathcal{M}_k(\mathcal{G})\}_{k=k_{\min}}^{k_{\max}}$ of orders $k_{\min}$ through $k_{\max}$ such that: (i) every $\mathcal{M}_k(\mathcal{G})$ is passive across all frequencies and Lyapunov stable; and (ii) its scattering response accurately approximates the true S-parameters of the structure.
\label{problem:main}
\end{problem}

\section{Proposed Framework}
\label{section:method}

Fig.~\ref{figure:overview} illustrates the proposed framework. A GNN encoder maps the layout graph $\mathcal{G}$ to an embedding $z$ (Section~\ref{section:graph}), from which parallel MLPs predict unconstrained raw blocks. Building upon a complete single-order pH parameterization (Section~\ref{section:head}), our nested formulation deterministically assembles a $k_{\max}$-state pH realization whose leading $k$ states directly form passive macromodels for all $k\in\{k_{\min},\dots,k_{\max}\}$ in one forward pass (Section~\ref{section:nested}). During training, the assembled models are evaluated at sampled frequencies and supervised directly on ground-truth S-parameters via an order-growth curriculum (Section~\ref{section:training}).

\subsection{Complete Single-Order pH Parameterization}
\label{section:head}

Neural networks naturally produce unconstrained real-valued outputs, whereas a valid $n$-state pH realization~\eqref{equation:ph} must satisfy the symmetry and definiteness conditions in~\eqref{equation:ph_cond}. To bridge this gap for a fixed order $n$ without sacrificing representational capacity, we construct a differentiable mapping from unconstrained raw blocks onto the complete set of admissible pH matrices.

The condition~\eqref{equation:ph_cond} imposes individual constraints on $J$, $N$, and $Q$, alongside a coupled positive semidefinite condition on $W$ (whose diagonal blocks $D_0$ and $R$ are themselves positive semidefinite). Leveraging standard skew-symmetric decompositions and triangular factorizations, we predict strictly lower-triangular blocks $\Lambda\in\mathbb{R}^{n\times n}$ and $\Gamma\in\mathbb{R}^{m\times m}$, lower-triangular blocks $L_D\in\mathbb{R}^{m\times m}$ and $L_R\in\mathbb{R}^{n\times n}$, and a lower-triangular block $L_Q\in\mathbb{R}^{n\times n}$ with softplus-activated positive diagonal entries, forming
\begin{equation}
\begin{aligned}
J&=\Lambda-\Lambda^{\mathsf{T}}, & N&=\Gamma-\Gamma^{\mathsf{T}}, & Q&=L_QL_Q^{\mathsf{T}},\\
D_0&=L_DL_D^{\mathsf{T}}, & R&=L_RL_R^{\mathsf{T}}.
\end{aligned}
\label{equation:factorsA}
\end{equation}
By skew-symmetric and Cholesky-type factorizations,~\eqref{equation:factorsA} spans all admissible $(J,N,Q,D_0,R)$.

To complete the dissipation matrix $W$, the off-diagonal port--state coupling $P$ cannot be omitted. While setting $P=0$ trivially ensures $W=\operatorname{diag}(D_0,R)\succeq0$, it forces $Z(s)-D$ to be positive real with $\tfrac{1}{2}(Z(j\omega)+Z(j\omega)^{\mathsf H})\succeq D_0$ at all frequencies. This excludes common passive structures whose resistance increases with frequency, such as an RL one-port $Z(s)=R_1+R_2\tau s/(1+\tau s)$ ($R_1, R_2,\tau>0$).

To retain all valid couplings, we use the contraction characterization of positive semidefinite block matrices~\cite{bhatia2007pd}: given $D_0=L_DL_D^{\mathsf{T}}$ and $R=L_RL_R^{\mathsf{T}}$, $W=\operatorname{diag}(L_D,L_R)\bigl[\begin{smallmatrix}I&K^{\mathsf{T}}\\K&I\end{smallmatrix}\bigr]\operatorname{diag}(L_D,L_R)^{\mathsf{T}}\succeq0$ if and only if $P=L_RKL_D^{\mathsf{T}}$ for some contraction $\|K\|_2\le1$. Thus, we predict an unconstrained $\tilde K\in\mathbb{R}^{n\times m}$ and normalize it as
\begin{equation}
P=L_RKL_D^{\mathsf{T}},\qquad K=\tilde K/\max\bigl(1,\|\tilde K\|_2\bigr),
\label{equation:coupling}
\end{equation}
which leaves contractions unchanged while guaranteeing $W\succeq0$ for arbitrary network weights.

\begin{figure*}[!htb]
\centering
\colorlet{fzThrSel}{fzkA}
\colorlet{fzThrGrey}{fzport}
\definecolor{fzThrRest}{HTML}{E8F0FA}
% ---------------------------------------------------------------------------------------------------------------------
% Glyph macros (prefix fzThr, local to this figure; from lens D1_twopart). u = mm per state, p = port block, n = state rows.
% Each matrix is its own glyph with its own brackets (\fzBrk); diagonals and frames as in \fzL, \fzLamone, \fzABCDkone.
% ---------------------------------------------------------------------------------------------------------------------
% L = [L_D 0; U T], origin = top-left corner of L (top of the port block). #1 u, #2 p, #3 n
\def\fzThrLopen#1#2#3{%
  \pgfmathsetmacro{\fzThrS}{#3*#1}%
  \fill[white] (0,0) rectangle ({#2+\fzThrS},{-#2-\fzThrS});
  \fill[fzThrGrey] (0,0) -- (0,-#2) -- (#2,-#2) -- cycle;}
% state rows #4+1 .. #5 of U and T filled with style #6. #1 u, #2 p, #3 n
\def\fzThrLband#1#2#3#4#5#6{%
  \pgfmathsetmacro{\fzThrS}{#3*#1}\pgfmathsetmacro{\fzThrYa}{-#2-#4*#1}\pgfmathsetmacro{\fzThrYb}{-#2-#5*#1}%
  \path[#6] (0,\fzThrYa) rectangle (#2,\fzThrYb);
  \begin{scope}\clip (#2,-#2) -- (#2,{-#2-\fzThrS}) -- ({#2+\fzThrS},{-#2-\fzThrS}) -- cycle;
    \path[#6] (#2,\fzThrYa) rectangle ({#2+\fzThrS},\fzThrYb);\end{scope}}
% partitions, diagonals of L_D and T, frame, brackets; #4 = 1: write the 0 of the zero block. #1 u, #2 p, #3 n
\def\fzThrLclose#1#2#3#4{%
  \pgfmathsetmacro{\fzThrS}{#3*#1}%
  \draw[fzhair] (0,-#2) -- ({#2+\fzThrS},-#2) (#2,0) -- (#2,{-#2-\fzThrS});
  \draw[fzinkii, line width=0.4pt] (0,0) -- (#2,-#2);
  \draw[fzoursdk, line width=0.55pt] (#2,-#2) -- ({#2+\fzThrS},{-#2-\fzThrS});
  \draw[fzmline] (0,0) rectangle ({#2+\fzThrS},{-#2-\fzThrS});
  \ifnum#4=1 \node[fzmath, text=fzinkii] at ({#2+\fzThrS/2},{-#2/2}) {$0$};\fi
  \fzBrk{0}{0}{#2+\fzThrS}{-#2-\fzThrS}}
% strictly lower-triangular Lambda (n x n), origin = top-left; zero diagonal dotted, 0.6 mm gap to the filled part. #1 u, #2 n
\def\fzThrLamopen#1#2{\pgfmathsetmacro{\fzThrS}{#2*#1}\fill[white] (0,0) rectangle (\fzThrS,-\fzThrS);}
% rows #3+1 .. #4 filled with style #5. #1 u, #2 n
\def\fzThrLamband#1#2#3#4#5{%
  \pgfmathsetmacro{\fzThrS}{#2*#1}%
  \begin{scope}\clip (0,-0.6) -- (0,-\fzThrS) -- ({\fzThrS-0.6},-\fzThrS) -- cycle;
    \path[#5] (0,{-#3*#1}) rectangle (\fzThrS,{-#4*#1});\end{scope}}
\def\fzThrLamclose#1#2{%
  \pgfmathsetmacro{\fzThrS}{#2*#1}%
  \draw[fzref, line width=0.5pt, dash pattern=on 0.5pt off 0.75pt, line cap=round] (0.15,-0.15) -- ({\fzThrS-0.15},{-\fzThrS+0.15});
  \draw[fzmline] (0,-0.6) -- ({\fzThrS-0.6},-\fzThrS);
  \draw[fzmline] (0,0) rectangle (\fzThrS,-\fzThrS);
  \fzBrk{0}{0}{\fzThrS}{-\fzThrS}}
% G (n x m), origin = top-left: rows #3+1 .. #4 filled with style #5. #1 u, #2 width
\def\fzThrGband#1#2#3#4#5{\path[#5] (0,{-#3*#1}) rectangle (#2,{-#4*#1});}
\def\fzThrGclose#1#2#3{\draw[fzmline] (0,0) rectangle (#2,{-#3*#1}); \fzBrk{0}{0}{#2}{-#3*#1}}
% realization (A,B,C,D) of the n-state model as ONE block matrix, origin = top-left; leading block at order k:
% A_k, B_k, C_k dark, the rest of A, B, C pale, D grey; dashed: boundary of the leading block (the cut at k).
% Block names written in the leading block. #1 u, #2 p, #3 n, #4 k
\def\fzThrABCD#1#2#3#4{%
  \pgfmathsetmacro{\fzThrS}{#3*#1}\pgfmathsetmacro{\fzThrK}{#4*#1}\pgfmathsetmacro{\fzThrW}{#3*#1+#2}%
  \fill[fzThrRest] (0,0) rectangle (\fzThrW,-\fzThrW);
  \fill[fzThrSel] (0,0) rectangle (\fzThrK,-\fzThrK);
  \fill[fzThrSel] (\fzThrS,0) rectangle (\fzThrW,-\fzThrK);
  \fill[fzThrSel] (0,-\fzThrS) rectangle (\fzThrK,-\fzThrW);
  \fill[fzThrGrey] (\fzThrS,-\fzThrS) rectangle (\fzThrW,-\fzThrW);
  \draw[fzpart] (\fzThrS,0) -- (\fzThrS,-\fzThrW) (0,-\fzThrS) -- (\fzThrW,-\fzThrS);
  \draw[fzmline, line width=0.5pt] (0,0) rectangle (\fzThrW,-\fzThrW);
  \draw[fzcut] (0,-\fzThrK) -| (\fzThrK,0)  (\fzThrS,-\fzThrK) -- (\fzThrW,-\fzThrK)  (\fzThrK,-\fzThrS) -- (\fzThrK,-\fzThrW);
  \node[fzmath] at ({\fzThrK/2},{-\fzThrK/2}) {$A_{\fzs{k}}$};
  \node[fzmath] at ({\fzThrS+#2/2},{-\fzThrK/2}) {$B_{\fzs{k}}$};
  \node[fzmath] at ({\fzThrK/2},{-\fzThrS-#2/2}) {$C_{\fzs{k}}$};
  \node[fzmath] at ({\fzThrS+#2/2},{-\fzThrS-#2/2}) {$D$};
  \fzBrk{0}{0}{\fzThrW}{-\fzThrW}}
% ---- part macros ----
\def\fzThrPartA{%
% L = [L_D 0; U T]: rows 1..k dark, rows k+1..k_max pale, L_D grey (shared)
\begin{scope}[shift={(5.4,5)}]
  \fzThrLopen{0.5}{5}{32}
  \fzThrLband{0.5}{5}{32}{0}{20}{fill=fzThrSel}
  \fzThrLband{0.5}{5}{32}{20}{32}{fill=fzThrRest}
  \fzThrLclose{0.5}{5}{32}{1}
\end{scope}
\node[fzmath, anchor=south] at (15.9,6) {$L$};
\node[fzmath, anchor=east] at (4,2.5) {$L_{\fzs{D}}$};
\node[fzmath] at (7.9,-13.9) {$U$};
\node[fzmath] at (13.12,-13.9) {$T$};
% Lambda (strictly lower triangular); name above the glyph, as L, G and (A,B,C,D)
\begin{scope}[shift={(30.4,0)}]
  \fzThrLamopen{0.5}{32}
  \fzThrLamband{0.5}{32}{0}{20}{fill=fzThrSel}
  \fzThrLamband{0.5}{32}{20}{32}{fill=fzThrRest}
  \fzThrLamclose{0.5}{32}
\end{scope}
\node[fzmath, anchor=base] at (38.4,1.8) {$\Lambda$};
% G (k_max x m)
\begin{scope}[shift={(50.4,0)}]
  \fzThrGband{0.5}{5}{0}{20}{fill=fzThrSel}
  \fzThrGband{0.5}{5}{20}{32}{fill=fzThrRest}
  \fzThrGclose{0.5}{5}{32}
\end{scope}
\node[fzmath, anchor=base] at (52.9,1.8) {$G$};
% the cut at k through every state-dependent output; the realization with its leading block
\draw[fzcut] (4.1,-10) -- (56.7,-10);
\node[fzmath, text=fzoursdk, anchor=west] at (57.2,-10) {$k$};
\draw[fzflow] (59.8,-10) -- (64.4,-10);
\begin{scope}[shift={(65.4,0)}] \fzThrABCD{0.5}{5}{32}{20} \end{scope}
\node[fzmath, anchor=base] at (75.9,1.8) {$(A,B,C,D)$};
% notes
\node[fznote, anchor=base, inner sep=0pt] at (25.9,-19.7) {rows \fzm{1,\dots,k};\ \fzm{L_{\fzs{D}}} shared};
\node[fznote, anchor=base east, inner sep=0pt] at (86.85,-24.4) {export = slicing, no refit};
}
\def\fzThrPartB{%
\begin{scope}[shift={(5.4,5)}]% stage 0
  \fzThrLopen{0.5}{5}{16}
  \fzThrLband{0.5}{5}{16}{0}{16}{fill=fzThrSel}
  \fzThrLclose{0.5}{5}{16}{0}
\end{scope}
\begin{scope}[shift={(28.9,5)}]% stage 1
  \fzThrLopen{0.5}{5}{24}
  \fzThrLband{0.5}{5}{24}{0}{16}{fill=fzThrSel}
  \fzThrLband{0.5}{5}{24}{16}{24}{fzbnew}
  \fzThrLclose{0.5}{5}{24}{0}
\end{scope}
\begin{scope}[shift={(56.4,5)}]% stage 2
  \fzThrLopen{0.5}{5}{32}
  \fzThrLband{0.5}{5}{32}{0}{24}{fill=fzThrSel}
  \fzThrLband{0.5}{5}{32}{24}{32}{fzbnew}
  \fzThrLclose{0.5}{5}{32}{0}
\end{scope}
\node[fzmath, anchor=east] at (4,-1.5) {$L$};% the stages are drawn for L
% state count of each stage over its glyph
\node[fznote, anchor=base, inner sep=0pt] at (11.9,6.8) {\fzm{k_{\fzs{0}}=k_{\fzs{\min}}}};
\node[fznote, anchor=base, inner sep=0pt] at (37.4,6.8) {\fzm{k_{\fzs{1}}}};
\node[fznote, anchor=base, inner sep=0pt] at (66.9,6.8) {\fzm{k_{\fzs{2}}=k_{\fzs{\max}}}};
% growth steps
\draw[fzflow] (19.85,-2) -- (27.45,-2);  \node[fznote, anchor=south] at (23.65,-1.25) {grow};
\draw[fzflow] (47.35,-2) -- (54.95,-2);  \node[fznote, anchor=south] at (51.15,-1.25) {grow};
% direct label of the appended (hatched) rows, left of them, last baseline on the bottom of the band
\node[fznote, anchor=base east, inner sep=0pt, align=right] at (27.4,-12) {\begin{tabular}[b]{@{}r@{}}new\\[-0.35ex]rows\end{tabular}};
\node[fznote, anchor=base east, inner sep=0pt, align=right] at (54.9,-16) {\begin{tabular}[b]{@{}r@{}}new\\[-0.35ex]rows\end{tabular}};
% order set of every training step of the stage: anchors k_0..k_g, then the random order kappa (blue)
\node[fznote, anchor=base, inner sep=0pt] at (11.9,-19.75) {train \fzm{\{k_{\fzs{0}}\}}};
\node[fznote, anchor=base, inner sep=0pt] at (37.4,-19.75) {train \fzm{\{k_{\fzs{0}},k_{\fzs{1}},\textcolor{fzoursdk}{\kappa}\}}};
\node[fznote, anchor=base east, inner sep=0pt] at (77.85,-19.75) {train \fzm{\{k_{\fzs{0}},k_{\fzs{1}},k_{\fzs{2}},\textcolor{fzoursdk}{\kappa}\}}};
}

\begin{tikzpicture}[fz, x=1mm, y=1mm]
\fzThrPartA
\node[font=\footnotesize\rmfamily, text=fzink, anchor=base, inner sep=0pt] at (43.15,-28.7) {(a) Nested pH parameterization};
\begin{scope}[shift={(102.45,-4.65)}]
\fzThrPartB
\end{scope}
\node[font=\footnotesize\rmfamily, text=fzink, anchor=base, inner sep=0pt] at (142.57,-28.7) {(b) Growth curriculum};
\end{tikzpicture}%
\caption{Multi-order state slicing and progressive training.
(a)~Nested pH parameterization: slicing the first $k$ rows of state-indexed blocks ($\Lambda,U,T,G$) while retaining the shared port blocks ($L_D,\Gamma$) yields a valid $k$-state passive realization $(A_k,B_k,C_k,D)$.
(b)~Order-growth curriculum (illustrated for $\bar{g}=2$): across stages $g=0,\dots,\bar{g}$ ($k_0=k_{\min}<\dots<k_{\bar{g}}=k_{\max}$), newly initialized state rows are progressively appended without altering lower-order models, jointly optimizing anchor orders and a randomly sampled intermediate order $\kappa$.}
\label{figure:nested_growth}
\end{figure*}
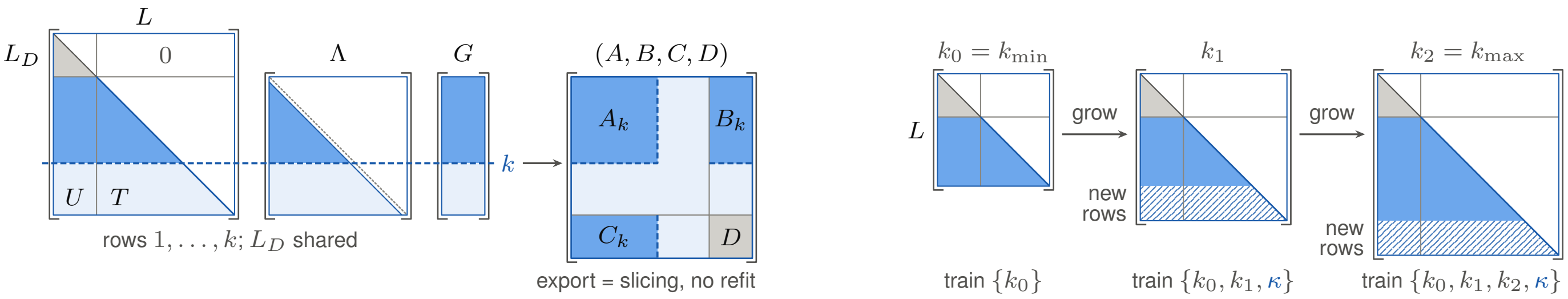

Together with an unconstrained $G\in\mathbb{R}^{n\times m}$,~\eqref{equation:ph} yields the $n$-state realization
\begin{equation}
\begin{aligned}
A&=(J-R)Q, & B&=G-P,\\
C&=(G+P)^{\mathsf{T}} Q, & D&=D_0+N.
\end{aligned}
\label{equation:realA}
\end{equation}

\begin{proposition}[Passivity and Completeness]
\label{proposition:passive}
(a)~For any unconstrained raw blocks,~\eqref{equation:factorsA}--\eqref{equation:realA} satisfies the pH conditions~\eqref{equation:ph}--\eqref{equation:ph_cond}, guaranteeing passivity across all frequencies and Lyapunov stability.
(b)~Conversely, every $n$-state pH realization~\eqref{equation:ph}--\eqref{equation:ph_cond} admits a parameterization of the form~\eqref{equation:factorsA}--\eqref{equation:realA}.
\end{proposition}
\begin{IEEEproof}
(a)~By~\eqref{equation:factorsA}, $J$ and $N$ are skew-symmetric, and $Q\succ0$ since $L_Q$ has positive diagonal entries. Because $\|K\|_2\le1$ in~\eqref{equation:coupling}, the block factorization of $W$ ensures $W\succeq0$. Thus~\eqref{equation:realA} satisfies~\eqref{equation:ph}--\eqref{equation:ph_cond}, which implies passivity and Lyapunov stability (Section~\ref{section:ph}).
(b)~Given any pH realization~\eqref{equation:ph}--\eqref{equation:ph_cond}, take $G$ unchanged, $\Lambda$ and $\Gamma$ as the strictly lower-triangular parts of $J$ and $N$, $L_Q$ the Cholesky factor of $Q$ (inverting softplus on its diagonal), and $L_D,L_R$ lower-triangular factors of $D_0,R$. Since $W\succeq0$, there exists a contraction $K$ satisfying $P=L_RKL_D^{\mathsf{T}}$~\cite{bhatia2007pd}. Therefore,~\eqref{equation:factorsA}--\eqref{equation:realA} recovers~\eqref{equation:ph}.
\end{IEEEproof}

Since Proposition~\ref{proposition:passive} holds for arbitrary MLP outputs, the network learns directly in the complete passive space without loss penalties or post-hoc passivity enforcement.

\subsection{Nested Multi-Order pH Parameterization}
\label{section:nested}

While Section~\ref{section:head} guarantees passivity for a fixed order $n$, Problem~\ref{problem:main} requires macromodels across orders $k \in \{k_{\min},\dots,k_{\max}\}$. Rather than training separate networks per order, we seek a single $k_{\max}$-state realization $(A,B,C,D)$ from which each $k$-state model can be extracted by slicing its first $k$ states, yielding the leading sub-realization $(A_k,B_k,C_k,D)$ with $A_k\in\mathbb{R}^{k\times k}$, $B_k\in\mathbb{R}^{k\times m}$, and $C_k\in\mathbb{R}^{m\times k}$. Specifically, we require two properties: (i)~\textit{prefix passivity}, ensuring every $(A_k,B_k,C_k,D)$ is passive and stable in one forward pass; and (ii)~\textit{prefix invariance}, keeping $(A_k,B_k,C_k,D)$ unchanged when states beyond $k$ are appended, enabling progressive state growth during training (Section~\ref{section:training}).

However, the single-order parameterization~\eqref{equation:factorsA}--\eqref{equation:realA} at $n=k_{\max}$ satisfies neither property. First, for general $Q\neq I$, leading principal submatrices of $A=(J-R)Q$ are not guaranteed to remain stable (e.g., for $J=\bigl[\begin{smallmatrix}0&1\\-1&0\end{smallmatrix}\bigr]$, $R=0.1I$, and $Q=\bigl[\begin{smallmatrix}1&0.9\\0.9&1\end{smallmatrix}\bigr]$, $A$ is Hurwitz with eigenvalues $-0.1\pm0.43j$ yet $A_{11}=+0.8>0$), violating prefix passivity. Second, the normalization in~\eqref{equation:coupling} takes the spectral norm $\|\tilde K\|_2$ over all $k_{\max}$ rows. Whenever $\|\tilde K\|_2>1$, appending states rescales the first $k$ rows of $P$ and breaks prefix invariance.

To resolve both issues, we fix the energy matrix to $Q=I$ and replace separate $D_0, R$ factors and spectral normalization with a single lower-triangular factorization of the joint dissipation matrix $W$, ordering the $m$-port block before the $k_{\max}$-state block. Specifically, the MLPs predict strictly lower-triangular $\Lambda\in\mathbb{R}^{k_{\max}\times k_{\max}}$ and $\Gamma\in\mathbb{R}^{m\times m}$, lower-triangular $L_D\in\mathbb{R}^{m\times m}$ and $T\in\mathbb{R}^{k_{\max}\times k_{\max}}$, and unconstrained $U,G\in\mathbb{R}^{k_{\max}\times m}$. With $J=\Lambda-\Lambda^{\mathsf{T}}$ and $N=\Gamma-\Gamma^{\mathsf{T}}$, we assemble
\begin{equation}
W=\begin{bmatrix}D_0&P^{\mathsf{T}}\\P&R\end{bmatrix}=LL^{\mathsf{T}},\qquad L=\begin{bmatrix}L_D&0\\U&T\end{bmatrix},
\label{equation:W}
\end{equation}
\begin{equation}
A=J-R,\quad B=G-P,\quad C=(G+P)^{\mathsf{T}},\quad D=D_0+N,
\label{equation:head}
\end{equation}
which expands to $D_0=L_DL_D^{\mathsf{T}}$, $P=UL_D^{\mathsf{T}}$, and $R=UU^{\mathsf{T}}+TT^{\mathsf{T}}$. This construction enforces $W\succeq0$ for arbitrary raw blocks without spectral normalization, while preserving the full set of representable passive transfer matrices, as established in Proposition~\ref{proposition:nested}(b).

Because $L$ is lower triangular with the port block ordered first, the leading blocks of $W$ are determined by the leading rows of $L$ alone. Specifically, the first $m+k$ rows of $L$ vanish beyond column $m+k$, yielding
\[
(LL^{\mathsf{T}})_{[1:m+k,\,1:m+k]}=L_kL_k^{\mathsf{T}},\qquad L_k=\begin{bmatrix}L_D&0\\U_k&T_k\end{bmatrix},
\]
where $U_k\in\mathbb{R}^{k\times m}$ denotes the first $k$ rows of $U$ and $T_k\in\mathbb{R}^{k\times k}$ the leading principal submatrix of $T$ (with $G_k\in\mathbb{R}^{k\times m}$ and $\Lambda_k\in\mathbb{R}^{k\times k}$ defined analogously). Since $\Lambda$ and $T$ are lower triangular, $(A_k,B_k,C_k,D)$ depends solely on the first $k$ rows of $\Lambda$, $U$, $T$, and $G$, while sharing the feedthrough $D$ across all orders (Fig.~\ref{figure:nested_growth}(a)).

\begin{proposition}[Prefix Nesting and Completeness]
\label{proposition:nested}
(a)~For any unconstrained raw blocks and every $k\in\{1,\dots,k_{\max}\}$, the leading sub-realization $(A_k,B_k,C_k,D)$ of~\eqref{equation:W}--\eqref{equation:head} is passive across all frequencies and Lyapunov stable (\textit{prefix passivity}), and depends only on $(\Lambda_k,U_k,T_k,G_k,L_D,\Gamma)$, remaining invariant to rows $k+1,\dots,k_{\max}$ (\textit{prefix invariance}).
(b)~For every $k\in\{1,\dots,k_{\max}\}$, any $k$-state pH realization~\eqref{equation:ph}--\eqref{equation:ph_cond} has the same transfer matrix $Z(s)$ as some $k$-state sub-realization $(A_k,B_k,C_k,D)$ of the form~\eqref{equation:W}--\eqref{equation:head}.
\end{proposition}
\begin{IEEEproof}
(a)~From~\eqref{equation:W}--\eqref{equation:head} and the block-triangular structure above, $(A_k,B_k,C_k,D)$ satisfies $A_k=J_k-R_k$, $B_k=G_k-P_k$, $C_k=(G_k+P_k)^{\mathsf{T}}$, and $D=D_0+N$, with skew-symmetric $J_k=\Lambda_k-\Lambda_k^{\mathsf{T}}$ and $\bigl[\begin{smallmatrix}D_0&P_k^{\mathsf{T}}\\P_k&R_k\end{smallmatrix}\bigr]=L_kL_k^{\mathsf{T}}\succeq0$. Thus, $(A_k,B_k,C_k,D)$ satisfies~\eqref{equation:ph}--\eqref{equation:ph_cond} with $Q=I_k$ and does not depend on rows $k+1,\dots,k_{\max}$.
(b)~Given any $k$-state pH realization~\eqref{equation:ph}--\eqref{equation:ph_cond} with $Q=L_QL_Q^{\mathsf{T}}\succ0$, applying the state similarity transformation $M=L_Q^{\mathsf{T}}$ preserves $Z(s)$ and yields $(\tilde A,\tilde B,\tilde C,D)=(L_Q^{\mathsf{T}} AL_Q^{-\mathsf T},\,L_Q^{\mathsf{T}} B,\,CL_Q^{-\mathsf T},\,D)$. Using $QL_Q^{-\mathsf T}=L_Q$, this simplifies to $\tilde A=\tilde J-\tilde R$, $\tilde B=\tilde G-\tilde P$, and $\tilde C=(\tilde G+\tilde P)^{\mathsf{T}}$, where $\tilde J=L_Q^{\mathsf{T}} JL_Q=-\tilde J^{\mathsf{T}}$, $\tilde G=L_Q^{\mathsf{T}} G$, $\tilde P=L_Q^{\mathsf{T}} P$, and $\tilde R=L_Q^{\mathsf{T}} RL_Q$. Moreover, $\tilde W=\bigl[\begin{smallmatrix}D_0&\tilde P^{\mathsf{T}}\\\tilde P&\tilde R\end{smallmatrix}\bigr]=\operatorname{diag}(I_m,L_Q^{\mathsf{T}})W\operatorname{diag}(I_m,L_Q)\succeq0$, which therefore admits a lower-triangular factorization $\tilde W=L_kL_k^{\mathsf{T}}$ with $L_k=\bigl[\begin{smallmatrix}L_D&0\\U_k&T_k\end{smallmatrix}\bigr]$. Taking $\Lambda_k,\Gamma$ as the strictly lower-triangular parts of $\tilde J,N$ and setting $G_k=\tilde G$ (with arbitrary entries in rows $>k$) recovers $(\tilde A,\tilde B,\tilde C,D)$, and hence $Z(s)$, via~\eqref{equation:W}--\eqref{equation:head}.
\end{IEEEproof}

By Proposition~\ref{proposition:nested}, passive macromodels of all orders $k\in\{k_{\min},\dots,k_{\max}\}$ can be exported from a single forward pass via direct matrix slicing, without post-hoc model order reduction~\cite{phillips2003prbt}.

\subsection{Order-Growth Curriculum on S-Parameters}
\label{section:training}

While~\eqref{equation:W}--\eqref{equation:head} guarantees passivity and prefix nesting structurally, training must ensure that every sliced sub-realization $(A_k,B_k,C_k,D)$ ($k_{\min}\le k\le k_{\max}$) accurately captures the physical port behavior. Because state-space realizations are non-unique up to similarity transformations, supervising $(A,B,C,D)$ directly would require pre-fitting and canonicalizing reference realizations. Instead, we supervise the network directly on observable S-parameters, requiring no rational fitting of any training structure. Specifically, for a training structure $i$ and order $k$, we evaluate the impedance $Z_k(j2\pi f_\ell)=C_k(j2\pi f_\ell I_k-A_k)^{-1}B_k+D$ at the sampled frequencies $\{f_\ell\}_{\ell=1}^{N_f}$ and apply the Cayley transform~\eqref{equation:cayley} to obtain the predicted scattering matrix $\hat{S}_k(f_\ell)$ (omitting structure index $i$ on matrices for brevity). Measuring every order directly against the ground-truth EM response $S(f_\ell)$, we define the per-structure error as
\begin{equation}
e_{i,k}=\Big(\tfrac{1}{N_f m^2}\sum_{\ell=1}^{N_f}\sum_{a,b=1}^{m} w_\ell\,\rho_\beta\big([\hat{S}_k(f_\ell)-S(f_\ell)]_{ab}\big)\Big)^{1/2},
\label{equation:loss}
\end{equation}
where $a,b$ index the ports, $\rho_\beta(z)=h_\beta(\operatorname{Re}z)+h_\beta(\operatorname{Im}z)$ applies the Huber function $h_\beta$ ($\beta=0.01$), and $w_\ell\propto 1+(f_\ell/f_{\max})^2$ (normalized to unit mean, with $f_{\max}=\max_\ell f_\ell$) mildly upweights the upper band to mitigate the high-frequency attenuation of state-space gradients.

Jointly optimizing multiple orders requires aggregating their mini-batch mean errors $\bar e_k$ over an active order set $\mathcal{K}\subseteq\{k_{\min},\dots,k_{\max}\}$, yet achievable approximation errors vary substantially with state count. Because an arithmetic average allows larger low-order residuals to dominate the gradient and hinder high-order refinement, we instead adopt a weighted geometric mean:
\begin{equation}
\mathcal{L}=\exp\Big(\sum_{k\in\mathcal{K}}\eta_k\log\bar e_k\Big),\qquad \sum_{k\in\mathcal{K}}\eta_k=1,
\label{equation:geomean}
\end{equation}
with weights $\eta_k>0$. Because its gradient satisfies $\nabla\mathcal{L}=\mathcal{L}\sum_{k\in\mathcal{K}}\eta_k(\nabla\bar e_k/\bar e_k)$, the update direction weights the relative sensitivity $\nabla\bar e_k/\bar e_k$ of each order by $\eta_k$, preventing coarse low-order errors from overshadowing high-order accuracy.

Exploiting the prefix invariance of Proposition~\ref{proposition:nested}(a), we optimize~\eqref{equation:geomean} through a $\bar g$-stage order-growth curriculum over state counts $k_0=k_{\min}<k_1<\dots<k_{\bar g}=k_{\max}$ (Fig.~\ref{figure:nested_growth}(b)), allowing higher-order realizations to build upon converged lower-order models. Specifically, stage $g=0,1,\dots,\bar g$ trains the network with $k_g$ states. Stage~0 trains the $k_{\min}$-state model alone, and each subsequent stage $g\ge 1$ appends $k_g-k_{g-1}$ state rows to $\Lambda$, $U$, $T$, and $G$ (initialized zero in $\Lambda$, near-zero in $U,G$, and diagonal in $T$). To avoid evaluating all $k_g-k_{\min}+1$ orders per step, stage $g\ge 1$ sets $\mathcal{K}=\mathcal{A}_g\cup\{\kappa\}$, pairing fixed anchor orders $\mathcal{A}_g=\{k_0,\dots,k_g\}$ with an order $\kappa$ sampled uniformly from $\{k_{\min}+1,\dots,k_g-1\}$ at each step.

\subsection{Layout Graph and Encoder}
\label{section:graph}

To predict the raw blocks $(\Lambda,\Gamma,L_D,U,T,G)$ across structures that vary in technology, routing topology, and primitive count, we represent each structure as a heterogeneous layout graph $\mathcal{G}$ paired with global stack-up parameters. Nodes in $\mathcal{G}$ represent ports and layout primitives (trace segments, vias, pads, reference patches, power/ground tiles, and decoupling elements), each carrying 15 numerical features (e.g., position, dimensions, and via/pad diameters) and 7 categorical features (e.g., primitive type, role, layer, and net group). Edges encode series connections, layer transitions, same-layer coupling, cross-layer proximity, and port or reference attachments, each with an edge type and 8 numerical features (e.g., spatial offsets, distance, overlap length, minimum gap, and reference flags). At the structure level, 70 numerical parameters (e.g., layout dimensions, layer thicknesses, $\varepsilon_r$, and $\tan\delta$) and 12 categorical identifiers (e.g., stack-up and material corner) specify the cross-section and materials. These global features are embedded and concatenated to each node input, edge input, and the pooled graph vector.

As shown in Fig.~\ref{figure:overview}, a GNN encoder maps $\mathcal{G}$ to a graph embedding $z\in\mathbb{R}^{512}$, and six parallel MLPs map $z$ to the raw blocks. Specifically, the encoder stacks three GATv2 layers~\cite{brody2022gatv2} (with 2, 2, and 1 attention heads of 256 channels), ELU activations, and layer normalization after the first two layers. Node features are aggregated via mean and max pooling and passed through a two-layer MLP with $\tanh$ activation, yielding a bounded embedding $z\in\mathbb{R}^{512}$ that prevents ill-conditioned initial matrix blocks. From $z$, six two-layer MLPs with 1024 hidden units and SiLU activations output the free entries of $\Lambda$, $\Gamma$, $L_D$, $U$, $T$, and $G$ in~\eqref{equation:W}--\eqref{equation:head}.

\section{Experiments}
\label{section:experiments}

\subsection{Experimental Setup}
\label{section:setup}

\paragraph{Dataset}
Our dataset comprises 270k 8-port blocks for modular package modeling across three physical families, each split 80/10/10 into training, validation, and test sets.
(1)~\textit{Die-to-die signal lanes} (120k structures, 85--142 graph nodes) comprise four sub-families: silicon interposer, silicon bridge, and two redistribution-layer variants, RDL-A and RDL-B (where lanes cross a bridged reference-plane slot).
(2)~\textit{Vertical transitions} (90k structures, 84--111 graph nodes) include through-silicon vias (TSVs), micro-bumps, and hybrid bonds.
(3)~\textit{Power delivery} (60k structures, 65 graph nodes) consists of power-mesh tiles on silicon and RDL interposers.

The ground-truth S-parameters are generated using Ansys HFSS 3D Layout 2024~R1~\cite{hfss} across $N_f=200$ frequencies: 0.25--50~GHz ($Z_0=50~\Omega$) for signal lanes and vertical transitions, and 0.1--20~GHz ($Z_0=1~\Omega$) for power delivery.

\begin{table*}[t]
\centering
\caption{Comparison of macromodel accuracy and passivity violations across nine sub-families (3{,}000 test structures each). \textit{RMS $|\Delta S|$}: mean / max error ($\times 10^{-3}$); \#violations: non-passive count at $\tau=10^{-6}$ / $10^{-4}$; \textit{Ours / Best}: mean-error ratio to the best baseline per row; bottom row: mean error/ratio and total violations.}
\label{table:main}
\scriptsize
\setlength{\tabcolsep}{3.2pt}
\begin{tabular*}{\textwidth}{@{\extracolsep{\fill}}ccc cc cc cc cc c@{}}
\toprule
\multirow{2}{*}[-0.5ex]{family} & \multirow{2}{*}[-0.5ex]{sub-family} & \multirow{2}{*}[-0.5ex]{$k$} & \multicolumn{2}{c}{MFT-NNLS} & \multicolumn{2}{c}{MATLAB RF Toolbox} & \multicolumn{2}{c}{Ansys Genequiv} & \multicolumn{2}{c}{Ours} & \multirow{2}{*}[-0.5ex]{\begin{tabular}{@{}c@{}}Ours /\\Best\end{tabular}} \\
\cmidrule(lr){4-5} \cmidrule(lr){6-7} \cmidrule(lr){8-9} \cmidrule(lr){10-11}
 & & & RMS $|\Delta S|$ & \#violations & RMS $|\Delta S|$ & \#violations & RMS $|\Delta S|$ & \#violations & RMS $|\Delta S|$ & \#violations & \\
\midrule
\multirow{4}{*}{\begin{tabular}{@{}c@{}}signal\\lanes\end{tabular}}
 & silicon interposer & 24 & 1.87 / 10.6 & 0 / 0 & 5.42 / 47.8 & 126 / 107 & 2.07 / 15.9 & 6 / 6 & \textbf{0.653} / \textbf{2.83} & \textbf{0 / 0} & 0.35 \\
 & silicon bridge & 24 & 0.964 / 4.30 & 0 / 0 & 3.06 / 59.1 & 52 / 33 & 1.27 / 4.88 & 0 / 0 & \textbf{0.489} / \textbf{2.03} & \textbf{0 / 0} & 0.51 \\
 & RDL-A & 32 & 3.20 / 17.2 & 37 / 33 & 6.49 / 94.2 & 47 / 42 & 2.88 / 11.8 & 3 / 2 & \textbf{1.20} / \textbf{3.53} & \textbf{0 / 0} & 0.42 \\
 & RDL-B & 32 & 2.92 / 19.9 & 4 / 4 & 6.86 / 121 & 350 / 308 & 2.91 / 12.1 & 5 / 0 & \textbf{1.33} / \textbf{4.35} & \textbf{0 / 0} & 0.46 \\
\midrule
\multirow{3}{*}{\begin{tabular}{@{}c@{}}vertical\\transitions\end{tabular}}
 & TSV & 24 & 1.19 / \textbf{7.50} & 0 / 0 & 1.56 / 57.2 & 41 / 35 & \textbf{0.925} / 12.7 & 0 / 0 & 1.04 / 10.7 & \textbf{0 / 0} & 1.12 \\
 & micro-bump & 16 & 0.634 / \textbf{0.899} & 0 / 0 & 1.33 / 5.94 & 42 / 1 & 1.46 / 8.50 & 10 / 10 & \textbf{0.363} / 1.86 & \textbf{0 / 0} & 0.57 \\
 & hybrid bond & 8 & 0.891 / 1.54 & 0 / 0 & 1.97 / 19.8 & 116 / 116 & 0.970 / 1.71 & 0 / 0 & \textbf{0.458} / \textbf{1.10} & \textbf{0 / 0} & 0.51 \\
\midrule
\multirow{2}{*}{\begin{tabular}{@{}c@{}}power\\delivery\end{tabular}}
 & silicon-interposer mesh & 24 & 1.30 / 3.78 & 0 / 0 & 2.67 / 7.08 & 0 / 0 & 1.74 / 10.4 & 7 / 5 & \textbf{1.01} / \textbf{3.22} & \textbf{0 / 0} & 0.78 \\
 & RDL-interposer mesh & 24 & 1.79 / 5.52 & 43 / 32 & 2.08 / 52.3 & 6 / 4 & 2.33 / 9.07 & 25 / 22 & \textbf{0.935} / \textbf{1.78} & \textbf{0 / 0} & 0.52 \\
\midrule
\multicolumn{3}{c}{average / total} & 1.64 / 7.92 & 84 / 69 & 3.49 / 51.6 & 780 / 646 & 1.84 / 9.67 & 56 / 45 & \textbf{0.831} / \textbf{3.48} & \textbf{0 / 0} & \textbf{0.58} \\
\bottomrule
\end{tabular*}
\end{table*}

\paragraph{Implementation and Training}
We train one unified network per family across all its sub-families using the nested parameterization~\eqref{equation:W}--\eqref{equation:head}. Each network is trained over $\bar g=2$ growth stages with uniform order weights $\eta_k$, setting anchor orders $(k_0,k_1,k_2)$ to $(16,24,32)$ for signal lanes and power delivery, and $(8,16,24)$ for vertical transitions. To guarantee strict Hurwitz stability alongside passivity, we shift $R \leftarrow R+\varepsilon I$ with $\varepsilon=10^{-3}$. Specifically, we train the base $k_0$-state model for $400$--$600$ epochs in stage~0 and grow the state space over two $200$--$400$-epoch stages. Models are optimized on NVIDIA RTX~6000~Ada GPUs using SOAP~\cite{vyas2024soap} with weight decay $0.03$ and EMA decay $0.9995$, setting initial learning rates to $10^{-3}$ in stage~0 and $1.5\times 10^{-4}$ during growth ($5\times$ higher for the $\Lambda$ MLP) and halving on validation plateaus to $10^{-5}$.

\paragraph{Baselines and Metrics}
We compare with three widely adopted macromodeling tools: MFT-NNLS~\cite{gustavsen2026mftnnls}, MATLAB RF Toolbox~\cite{rationalfit}, and Ansys Genequiv~\cite{genequiv}, all fitted to ground-truth EM responses whereas our networks predict directly from unseen layouts. Since $p$ common poles across $m=8$ ports yield $mp$ states (only $k/m$ poles in a direct $k$-state fit), for each target order $k$ we sweep $p = \lceil k/m\rceil, \lceil k/m\rceil+1, \dots$, apply built-in passivity enforcement and bounded-real balanced truncation~\cite{opdenacker1988brbt} to $k$ states (or standard balanced truncation~\cite{moore1981bt} when residual non-passivity precludes it), stop when mean test error improves by under $1\%$, and report the best $p$. Prior S-parameter surrogates~\cite{traceformer,pcbformer,liu2024gnnsp,zhang2022pdncnn} and neuro-TF models~\cite{feng2016neurotf,xu2024deepotf} are omitted: fitting exact EM responses already provides an accuracy upper bound on surrogate-plus-fitting flows, whereas neuro-TF models apply only to fixed-topology design parameters. Moreover, neither guarantees continuous-frequency passivity without post-hoc fitting or enforcement ($0.1$--$7.7$~s).

We evaluate the frequency-domain accuracy via the unweighted root-mean-square error $\text{RMS }|\Delta S| = \big(\frac{1}{N_f m^2}\sum_{\ell,a,b}|\hat S_{ab}(f_\ell)-S_{ab}(f_\ell)|^2\big)^{1/2}$ against ground-truth EM labels, and verify full-spectrum passivity across $\omega\in[0,\infty]$, flagging a model as non-passive at tolerance $\tau \in \{10^{-6}, 10^{-4}\}$ if it is unstable or satisfies $\sup_{\omega\in[0,\infty]}\sigma_{\max}(S(j\omega)) > 1+\tau$.

\subsection{Main Results}
\label{section:main_results}

\paragraph{Accuracy and Passivity}
Table~\ref{table:main} compares macromodel accuracy and passivity across all nine sub-families. Because electrical complexity varies widely across sub-families (from short hybrid bonds to slotted RDL channels), each sub-family is evaluated at the smallest anchor order $k$ where the most accurate baseline attains a validation mean RMS $|\Delta S| \le 2\times 10^{-3}$, capped at $k_{\max}$.

Our networks reduce mean RMS $|\Delta S|$ by $42\%$ on average over each row's strongest baseline ($49\%$--$76\%$ in overall mean against each individual tool), achieving the lowest mean error in eight of nine sub-families and the lowest maximum error in seven, despite predicting directly from unseen layout graphs without test-time EM simulation. Specifically, our sliced models lower mean error by $49\%$--$65\%$ on signal lanes, $43\%$--$49\%$ on micro-bumps and hybrid bonds, and $22\%$--$48\%$ on power-delivery meshes; on TSV, our model outperforms MFT-NNLS and RF Toolbox while coming within $12\%$ of Ansys's realization with lower maximum error ($10.7$ vs.\ $12.7$). Whereas baselines fit poles per structure followed by post-hoc perturbation and truncation, our networks optimize directly within the passive space.

For passivity, our networks achieve zero violations ($0/0$) across all sub-families. Moreover, all $459{,}000$ realizations exported across all 17 orders ($k=k_{\min},\dots,k_{\max}$) of the 27k test structures remain stable and satisfy $\sigma_{\max}(S(j\omega)) \le 1$ up to machine precision over $\omega\in[0,\infty]$. In contrast, the baselines leave $56$--$780$ non-passive models in total ($\tau=10^{-6}$, Table~\ref{table:main}) even after built-in passivity enforcement. These violations often peak between or above the sampled frequencies where discrete-grid checks miss them, and even small violations can trigger instability under passive terminations.

\begin{figure}[!htb]
\centering
\includegraphics[width=0.7\columnwidth]{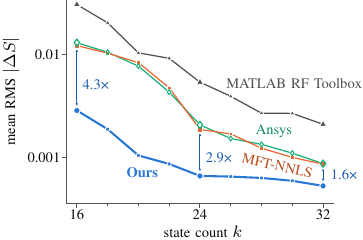}
\caption{Multi-order test accuracy on silicon-interposer lanes. The arrows show the mean-error gap and ratio to the best baseline.}
\label{figure:order}
\end{figure}

\paragraph{Multi-Order Slicing}
Fig.~\ref{figure:order} plots test error on silicon-interposer lanes from $k=16$ to $32$. Whereas baselines require over-parameterized fitting, passivity enforcement, and balanced truncation per target order, one forward pass of our network yields sliced realizations that smoothly reduce error across all orders and outperform all three tools, with $k=24$ ($6.53\times 10^{-4}$) already surpassing all 32-state baseline macromodels. While $k=32$ still yields $54\%$--$58\%$ reductions on complex RDL lanes (Table~\ref{table:main}), the curve here flattens above $k=24$ as residuals approach the single-solve EM noise floor, where further raw-label reduction increasingly reflects fitting solver jitter (Section~\ref{subsection:discussion}).

\paragraph{Runtime}
Table~\ref{table:time} compares per-structure macromodel generation time on a single Xeon Gold 6430 CPU core. Whereas conventional flows require $4.9$--$77$ core-minutes per structure for full-wave EM simulation (over an hour for half of RDL lanes) plus $0.1$--$7.7$~s for rational fitting and passivity enforcement, our framework maps a raw layout to its full-order $(A,B,C,D)$ realization (and all lower-order slices) in a median of $36$--$85$~ms, achieving an $11{,}000$--$39{,}000\times$ speedup and enabling sub-$100$~ms turnaround for iterative package layout exploration once trained per technology family.

\begin{table}[!htb]
\centering
\caption{Per-structure macromodel generation time on a single Xeon Gold 6430 CPU core. EM Solve reports mean (range across sub-families); Ours reports median layout-to-realization time.}
\label{table:time}
\scriptsize
\setlength{\tabcolsep}{5pt}
\begin{tabular}{@{}cccc@{}}
\toprule
family & EM Solve (core-min) & Ours (ms) & speedup \\
\midrule
signal lanes & 54.9 (32--77) & 85 & 39{,}000$\times$ \\
vertical transitions & 12.5 (4.9--21) & 69 & 11{,}000$\times$ \\
power delivery & 18.6 (15--22) & 36 & 31{,}000$\times$ \\
\bottomrule
\end{tabular}
\end{table}

\subsection{Robustness to EM Solver Noise}
\label{subsection:discussion}

Beyond improving accuracy on standard EM labels, cross-layout passive learning from deterministic geometry confers robustness to numerical fluctuations in full-wave solvers. Specifically, on a randomly sampled subset of signal-lane test structures, we average repeated HFSS solves per structure under perturbed meshing and sweep settings to form a denoised S-parameter reference. Re-evaluating the existing macromodels against this independent reference without retraining or refitting reveals opposite trends. The mean RMS $|\Delta S|$ for our layout-predicted models drops from $7.8\times 10^{-4}$ (against the original labels) to $5.6\times 10^{-4}$ (median per-structure ratio $0.66\times$). In contrast, the mean error rises from $1.27\times 10^{-3}$ to $1.50\times 10^{-3}$ (median $1.20\times$) for MFT-NNLS, from $1.39\times 10^{-3}$ to $1.65\times 10^{-3}$ ($1.15\times$) for Ansys Genequiv, and from $4.0\times 10^{-3}$ to $4.1\times 10^{-3}$ ($1.03\times$, largely masked by higher baseline error) for MATLAB RF Toolbox. Because per-structure fitters optimize against a single EM solve, their poles and residues partially absorb run-specific solver jitter once model fidelity approaches the label noise floor. Conversely, our network maps deterministic layouts to pH realizations across tens of thousands of training structures, averaging out uncorrelated solver noise during training while remaining insulated from test-label fluctuations at inference. This contrast highlights an implicit denoising benefit of cross-layout learning, which captures the underlying physical response rather than single-solve numerical artifacts.

\section{Conclusion}
\label{section:conclusion}

This paper presented an end-to-end macromodeling framework that maps physical layouts directly to nested, continuous-frequency passive port-Hamiltonian macromodels.
Experiments on 27k unseen test structures from a 270k-structure 8-port advanced-packaging dataset confirm $42\%$ average ($65\%$ peak) lower mean error than established fitting tools, zero passivity violations, and sub-$100$~ms macromodel generation ($11{,}000$--$39{,}000\times$ speedup).

\bibliographystyle{IEEEtran}
\bibliography{refs}

\end{document}